\documentclass[%
aip,
amsmath,amssymb,
reprint,nofootinbib%
]{revtex4-2}
\usepackage{nomencl}
\makenomenclature
\renewcommand\nomgroup[1]{%
  \item[\bfseries
  \ifstrequal{#1}{A}{}{%
  \ifstrequal{#1}{N}{Abbreviations}{%
  \ifstrequal{#1}{B}{Subscripts}{%
  \ifstrequal{#1}{C}{Superscripts}{%
  \ifstrequal{#1}{D}{Greek Letters}{%
  \ifstrequal{#1}{X}{Other symbols}{}}}}}}%
]}

\usepackage{graphicx,array,booktabs,rotating,parskip}
\usepackage{dcolumn}
\usepackage{bm}
\usepackage{mathptmx}
\usepackage{comment}
\usepackage{enumitem}
\usepackage{afterpage,float,color,xcolor}
\usepackage{multirow}
\usepackage{lipsum} 
\usepackage{hyperref}
\usepackage{etoolbox} 
\usepackage{tabularx,tabulary}
\usepackage[capitalize]{cleveref}
\usepackage{times}
\hypersetup{
	colorlinks,
	linkcolor={blue!100!black},
	citecolor={blue!100!black},
	urlcolor={blue!100!black}
}

\makeatletter
\newcommand\footnoteref[1]{\protected@xdef\@thefnmark{\ref{#1}}\@footnotemark}
\makeatother

\newcommand{\etal}{\textit{et al.}}
\usepackage[page]{totalcount}
\usepackage{etoolbox,fancyhdr,xcolor}%
\usepackage{ragged2e}
\usepackage[displaymath,pagewise]{lineno}
\usepackage{tikz}
\newcommand*\circled[1]{\tikz[baseline=(char.base)]{
            \node[shape=circle,draw,inner sep=0.5pt] (char) {#1};}}

\begin{document}
\preprint{AIP/123-QED}
\title[\textbf{Journal to be decided} (2026) $|$  Rough draft of the manuscript]{Axisymmetric cavities in hypersonic flow}

\author{Soumya R. Nanda}
\email{soumyananda224@gmail.com}
\affiliation{Department of Aerospace Engineering, Indian Institute of Technology Kanpur, Kanpur-208016, Uttar Pradesh, India}

\author{Talluri Vamsi Krishna}
\email{sri.vamsi1432@gmail.com}
\affiliation{Faculty of Aerospace Engineering, Technion-Israel Institute  of Technology, Haifa-3200003, Israel}

\author{S. K. Karthick}
\email{skkarthick@mae.iith.ac.in (corresponding author)}
\affiliation{Department of Mechanical and Aerospace Engineering, Indian Institute of Technology Hyderabad, Kandi-502285, Telangana, India}

\author{Jacob Cohen}
\email{aerycyc@technion.ac.il}
\affiliation{Faculty of Aerospace Engineering, Technion-Israel Institute  of Technology, Haifa-3200003, Israel}

\date{\today}

\begin{abstract}

A detailed experimental campaign is conducted to assess the shear layer characteristics of an axisymmetric open cavity configuration exposed to a Mach $6$ freestream flow. Experiments are carried out in a Ludwieg tunnel across varying Reynolds number cases ($23000\leq Re_D \leq 74000$) based on the cavity depth ($D$). The influence of associated geometrical parameters in the form of length-to-depth ratio ($[L/D]=[2,4,6]$) and non-dimensional excess height of the rear face with respect to the front face ($[\Delta h/D] = [-0.5,-0.25,0,0.25,0.5]$) is assessed. Thereby, a detailed interpretation of the shear layer evolution is made based on qualitative flow structures captured using schlieren imaging and Planar Laser Rayleigh Scattering (PLRS), along with quantitative measures obtained through unsteady pressure probes. Regardless of the $[L/D]$, the state of the shear layer is observed to be laminar at lower $Re_D$s, and when increased, reveals the presence of Kelvin-Helmholtz (K-H) vortices. Nevertheless, for the longest aspect-ratio case ($[L/D]=6.0$), the shear layer transitions to turbulence in the extreme $Re_D$ cases owing to the availability of a longer length scale for K-H propagation. The spectral content obtained from the pressure response and the light intensity from the PLRS snapshots reveals an increase in the dominant frequency from the first Rossiter mode to higher-order modes for $[L/D]=6.0$. Apart from $[L/D]=6, [\Delta h/D]=0$, the dominant frequency for all the cases considered herein matches the prediction of Rossiter modes and is found to be invariant of Reynolds number. While varying the $[\Delta h/D]$, switching of modes occurs as obtained using Proper Orthogonal Decomposition (POD) analysis of PLRS snapshots. For the negative $[\Delta h/D]$, the K-H mode is dominant (5th-6th Rossiter modes), whereas for the positive, the strong flapping mode (1st Rossiter mode) prevails owing to significant pressure build-up inside the cavity recirculation region. For $[\Delta h/D]=0$, both modes may co-exist depending on $Re_D$. The azimuthal uniformity of the flow is also probed, indicating the dominance of the axisymmetric mode in the flapping scenario and much less correlated behaviour in the case of K-H vortices in the shear layer.    
\end{abstract}
 
\maketitle
\nomenclature[A]{$L$}{Cavity length (m)}
\nomenclature[A]{$w$}{Cavity width (m)}
\nomenclature[A]{$L_F$}{Upstream cavity length from the cone tip (m)}
\nomenclature[A]{$D$}{Cavity depth (m)}
\nomenclature[A]{$p$}{Instantaneous static pressure (Pa)}
\nomenclature[A]{$T$}{Instantaneous static temperature (K)}
\nomenclature[A]{$M$}{Mach number (-)}
\nomenclature[A]{$u$}{Instantaneous velocity (m/s)}
\nomenclature[A]{$Re$}{Reynolds number per m ($\rho u /\mu$, 1/m)}
\nomenclature[A]{$f$}{Frequency (Hz)}
\nomenclature[A]{$n$}{Mode number (-)}
\nomenclature[A]{$St$}{Strouhal number (-)}
\nomenclature[A]{$\Delta h$}{Excess rear-face height (m)}
\nomenclature[A]{$s$}{Streamwise distance along the shear layer (m)}
\nomenclature[A,3]{}{}
\nomenclature[A]{}{}

\nomenclature[B]{$\infty$}{Freestream quantity}
\nomenclature[B]{$0$}{Total or stagnation quantity}
\nomenclature[B]{$2$}{Quantities incoming to the cavity}
\nomenclature[B]{$e$}{Nozzle exit}
\nomenclature[B]{$f$}{front face}
\nomenclature[B]{$r$}{rear face}
\nomenclature[B]{$D$}{Based on cavity depth}
\nomenclature[B,3]{}{}
\nomenclature[B]{}{}

\nomenclature[C]{$\overline{\Box}$}{Time-averaged quantity}
\nomenclature[C]{$\tilde{\Box}$}{Re-scaled quantity}
\nomenclature[C,3]{}{}
\nomenclature[C]{}{}

\nomenclature[D]{$\rho$}{Density (kg/m$^3$)}
\nomenclature[D]{$\mu$}{Dynamic viscosity (kg/ms)}
\nomenclature[D]{$\zeta$}{Normalized pressure}
\nomenclature[D]{$\theta$}{Semi-apex angle ($^\circ$)}
\nomenclature[D]{$\beta$}{Leading edge shock angle ($^\circ$)}
\nomenclature[D]{$\lambda_i$}{Eigen values}
\nomenclature[D]{$\sigma_i$}{Energy fraction of $i$th POD mode}
\nomenclature[D,3]{}{}
\nomenclature[D]{}{}

\nomenclature[N]{POD}{Proper Orthogonal Decomposition}
\nomenclature[N]{FFT}{Fast Fourier Transform}
\nomenclature[N]{C-D}{Convergent-Divergent}
\nomenclature[N]{K-H}{Kelvin-Helmholtz}
\nomenclature[N]{PLRS}{Planar Laser Rayleigh Scattering}
\nomenclature[N,3]{}{}
\nomenclature[N]{}{}

\printnomenclature


\section{Introduction} \label{sec:intro}

Cavities in high-speed flows have crucial aspects owing to the complex interaction of associated shock waves, separated shear layers, and the recirculation region \citep{Beresh_Wagner_Casper_2016}. Cavities find applications in various areas, including landing gear housings \citep{Guo}, weapon bays \citep{Robertson,Lawson}, scramjet components\citep{kang2011cowl}, flow control \citep{silton2005use}, and flame stabilization \citep{ben1998investigation,CAO2021}, among others. The cavity dynamics, specifically shock-shear layer interaction and recirculation bubble characteristics, vary significantly due to compressibility effects as the Mach number is changed, resulting in an unsteady flowfield in most cases, accompanied by spanwise coherent vortices \citep{liu2021, murray2001}. The current investigation focuses on hypersonic flow past an axisymmetric cavity, which often finds application in both small- and large-scale contexts, such as sensor housing provisions\citep{mcgilvray2009helmholtz} and trapped combustor\citep{mathur2001supersonic}. Nevertheless, the applicability point of view differs in both cases. Small-scale cavities are used as a thermal shielding mechanism to isolate the sensors from the elevated heating zone, providing a laminar separated shear layer that results in reduced heat transfer\citep{chapman1958}. Large-scale cavities, on the other hand, are typically used to aid rapid flow mixing. Moreover, upon reattachment of the separated flowfield, a hotspot forms\citep{nicoll1964} in the cavity's trailing edge. The transition from a laminar to a turbulent state in the incoming shear layer results in a further increment in the trailing edge heat load\citep{Everhart2010}. Therefore, it is essential to comprehend the state of the shear layer, its characteristics, and potential interaction mechanisms with the cavity's trailing edge, which may trigger unsteadiness.\\ 

The overall cavity flow is typically governed by the feedback loop of shear layer impingement on the aft wall, along with the generation and upstream propagation of acoustic waves, which then perturb the shear layer, thereby completing the feedback cycle \citep{morgenstern1994}. The perturbations introduce vortical structures in the shear layer through K-H instability, which, when in a proper phase with the acoustic waves, results in a strong resonance leading to self-sustained oscillations, referred to as Rossiter modes \citep{rossiter1964, rowley2002, sun2017}. The incoming flow properties and the geometrical parameters concerning the cavity are among the key attributes that drive the overall flow oscillation. Rossiter\citep{rossiter1964}, followed by Heller \etal~\cite{Heller1971}, proposed a semi-empirical relation to predict the associated resonance frequency, which is presented in terms of the Strouhal number ($St_D$) as,
\begin{equation}
\label{eq:rossiter}
    St_D = \left[\frac{fD}{u}\right] = \frac{(n-\alpha)(D/L)}{M\left(1+\left[(\gamma-1)/2\right]M^2\right)^{-1/2}+1/k},
\end{equation}

where $M$ represents the Mach number of the flow upstream of the cavity, and the integer number $n = [1, 2, 3, 4, \ldots]$ indicates the mode number. The empirical constants $\alpha = 0.25$ and $k = 0.57$ \citep{murray2001} correspond to the phase lag of the feedback loop, and the velocity ratio of the K-H convection to the freestream, respectively. Amplification or stabilization of the Rossiter modes is typically reported for varying geometric situations (cavities with negative and positive asymmetry\citep{rezende2024}) and incoming flow compressibility \citep{mathias2022}. In contrast, cavity flows in relatively higher supersonic flow regimes are observed to be driven by the contribution of additional waves in excess of the impingement of the trailing-edge vortices \citep{li2013}. The additional wave structures arise from the differences in the relative speed of the convecting large-scale structures with respect to the freestream flow and the internal cavity flow \citep{papamoschou1988}. An increase in the Mach number also led to the dominance of three-dimensional (3D) modes over two-dimensional (2D) modes \citep{bres2008, sandham1991, doshi2022}.\\

Two-dimensional cavities are relevant to real-world applications and can also serve as a canonical model to capture complex fields. Several researchers have investigated such cavities across various Mach numbers, Reynolds numbers, and cavity geometries. Recently, modal and resolvent analysis have helped establish the K-H growth rates \citep{garrido2014} and cross-frequency interactions involving energy transfer from the fundamental Rossiter mode to harmonic modes \citep{islam2024}, thereby identifying the flow behavior as convectively amplified or highly oscillatory. The stability analysis \citep{sun2017, de2014} is reported to demonstrate the significance of 2D Rossiter modes, along with 3D eigenmodes, including centrifugal instabilities and spanwise structures, in longer cavities. The evolution of tornado-like structures \citep{luo2023} and suppression of turbulence intensity with increase in Mach number \citep{Beresh_Wagner_Casper_2016} also happens to be altering the dominant spectral content through disruption of coherence in Rossiter modes. Most of the available literature emphasizes the interplay between compressibility effects, modal instabilities, and three-dimensional structures in the evolution of flow in two-dimensional cavities \citep{aguirre2025}.

\begin{figure*}
\includegraphics[width=0.8\textwidth]{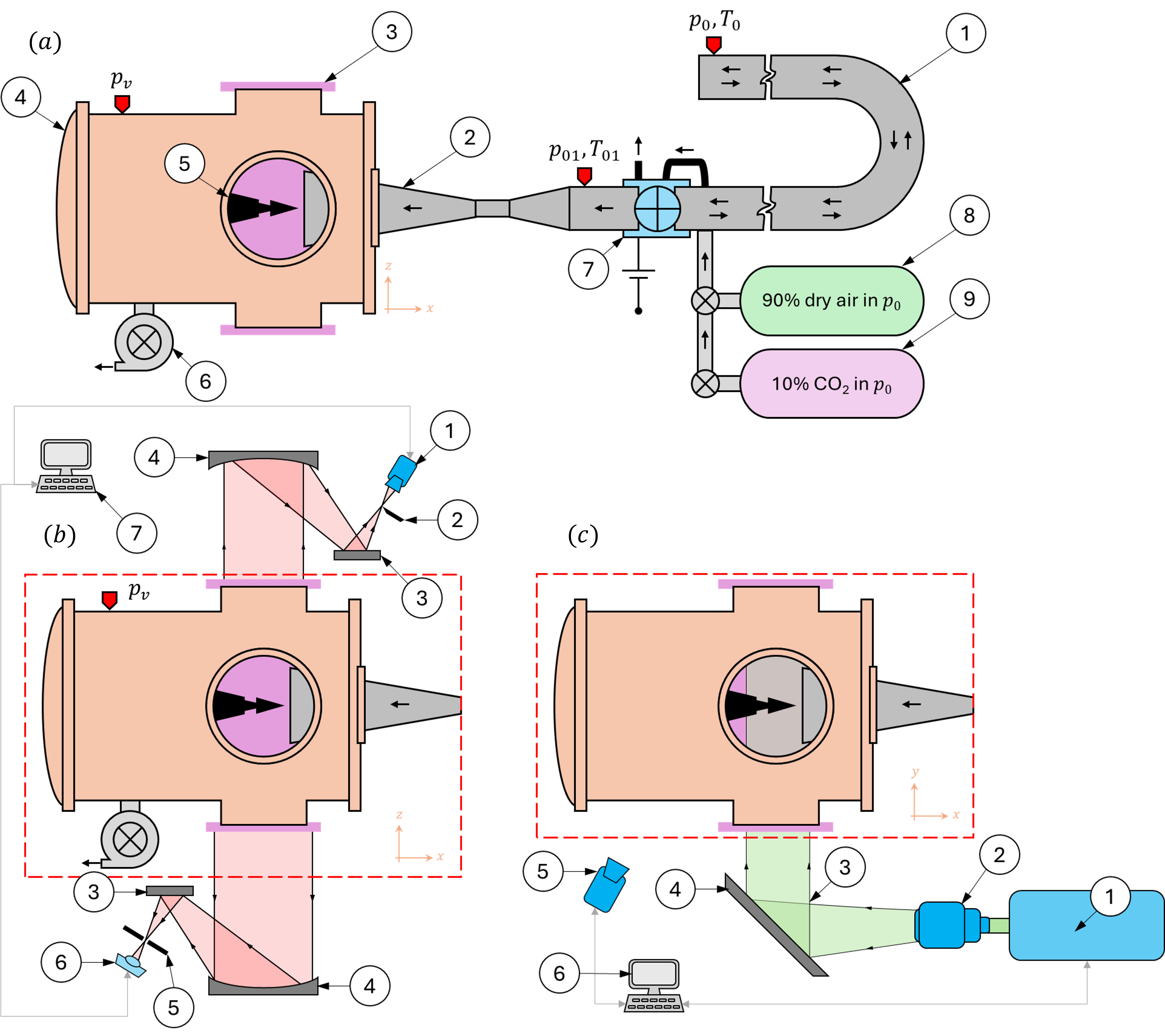}
\caption{(a) A typical schematic showing the short-duration hypersonic Ludwieg tunnel at Technion. Key components: 1. driver tube, 2. converging-diverging (C-D) nozzle, 3. view/optical ports, 4. dump/vacuum tank, 5. model mounting station, 6. vacuum pump, 7. fast-acting valve, 8. dry-air cylinder, and 9. CO$_2$ cylinder; (b) A schematic showing the high-speed schlieren imaging setup. Key components: 1. high-speed camera, 2. knife edge, 3. plane mirror, 4. parabolic mirror, 5. slit, 6. light source, and 7. computer; (c) A schematic showing the high-speed planar laser Rayleigh scattering setup. Key components: 1. continuous wave laser (532 nm), 2. sheet optics + collimator, 3. thin laser sheet, 4. plane mirror, 5. high-speed camera, and 6. computer.}
\label{fig:facility}
\end{figure*}

Although 2D cavity configurations have been extensively investigated across various flow regimes, their axisymmetric counterparts appear to be less explored. In the case of 2D planar cavities, where the spanwise coherence of vortical structures plays a crucial role \citep{casper2018}, acoustic wave propagation is primarily unidirectional due to the coupling between vortex shedding and pressure resonance. Nevertheless, in axisymmetric cavities, information can propagate in the azimuthal direction, potentially altering the cavity dynamics as a whole. Recently, due to the growing interest in developing hypersonic propulsion systems\citep{haberle2008investigation} and redistributing thermal loads on reentry vehicles\citep{yuceil1995nose}, the axisymmetric cavity has emerged as a more practical approach. Jackson \etal~\cite{jackson2001} investigated annular cavity configurations at varying aspect ratios for an incoming Mach number of $8.9$ and reported a laminar nature of the shear layer for the shorter cavities, and transitional for $[L/D]=8$. For longer cavities, three-dimensional unsteadiness is observed, likely due to the formation of a Taylor-Gortler instability. Similar research was undertaken by Creighton and Hillier (2007) \cite{creighton2007}, where different aspect-ratio cavities were mounted on a cone and exposed to an incoming flow with a Mach number of 5.5. The observed shear layer motion is classified as steady, mildly unsteady, and highly unsteady for different $[L/D]$ cases in conjunction with variation in the ratio between the upstream cavity length ($L_F$) and the length of the cavity ($L_C$). Based on that, the following criterion was deduced to label the flow unsteadiness, 
\begin{equation}
\label{eq:unsteadiness_identify}
    \frac{L_C}{L_F} \approx 2\left[\frac{\tan(\beta-\theta)}{\tan(\mu)-\tan(\beta-\theta)} \right],
\end{equation}
where $\theta,\ \beta,\ \mu$ are the cone angle, leading edge shock wave angle, and Mach angle, respectively.  Additionally, it was noted that for the highly oscillatory shear layer case, the axisymmetry of the flow is preserved. In contrast, for the mild unsteady cases, a poor correlation is observed between the symmetric pressure sensors. The nature of the unsteady shear layer, in particular, has a potential that can manifest complex wave propagation within the cavity and result in the shifting of Rossiter modes. Interestingly, the Rossiter mode, which enables the identification of resonant frequencies in cavities, is also independent of the incoming flow Reynolds number\citep{block1976noise}.  

Most of the existing cone-cavity studies do not account for the effect of the Reynolds number on the shear layer characteristics and acoustic feedback mechanism. In this context, the present study aims to systematically investigate the characteristics of the shear layer and its associated unsteadiness. An axisymmetric cavity mounted on a cone with a semi-apex angle of $24^\circ$ and a provision for varying the length-to-depth ratio as $ [L/D] = [2, 4, 6]$ is considered for conducting a detailed experimental campaign. The influence of the reattachment surface is particularly explored by varying the normalized excess height of the rear cavity face as $[\Delta h/D] = [-0.25,0,0.5]$. These geometric variations of the cavity, along with a wide range of Reynolds number cases ($23000 \leq Re_D \leq 74000$), are investigated for a freestream design Mach number of $M_\infty = 6$. Qualitative methods, such as schlieren and Rayleigh scattering-based imaging, are used in conjunction with quantitative measurements from unsteady pressure transducers to characterize the overall flowfield. This enables us to better understand the self-sustained oscillation pattern, along with the possible mechanism of driving the unsteadiness in axisymmetric cavity flows.


\section{Experimental Methodology} \label{sec:exp_meth}

\subsection{Ludwieg Tunnel Facility Description} \label{ssec:ludwieg}

\begin{figure*}
\includegraphics[width=0.85\textwidth]{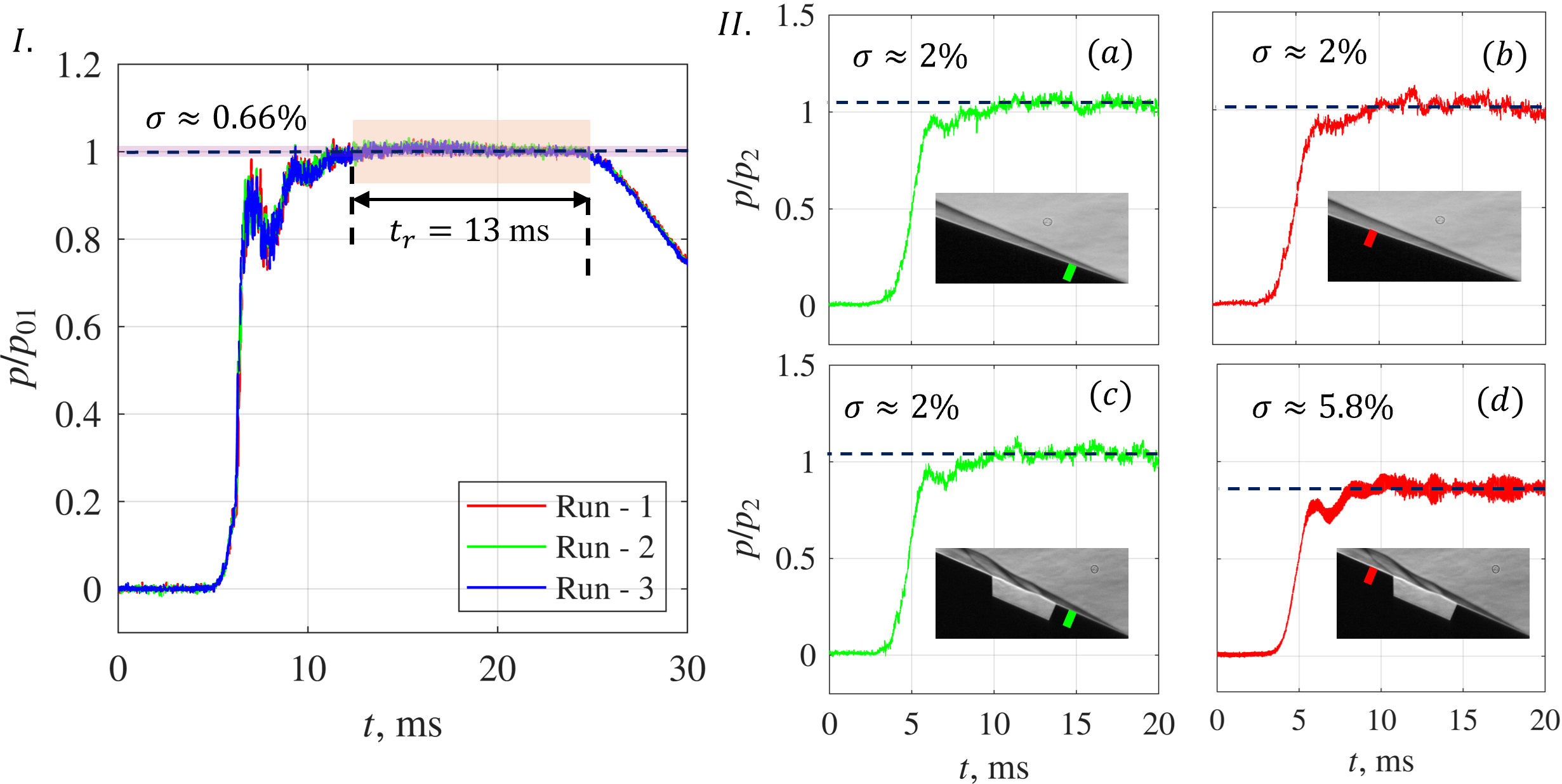}
\caption{ (I) Typical pressure plot depicting the repeatability of the signal obtained from a location just upstream of the C-D nozzle. The fluctuation intensity ($\sigma$) across the runs for $p/p_{01}$ is found to be $\approx 0.66\%$. (II) The local pressure depicting the fluctuation levels at the indicated locations marked in the snippet for the plain cone (a,b), and the cavity-mounted cone (c,d). $p_2$ corresponds to the static pressure post the conical leading edge shock, as can be referred from Table \ref{tab:freestream_cond}. $t_r$ refers to the run time for which the incoming pressure is found to be approximately constant.}
\label{fig:storage press}
\end{figure*}

The Ludwieg tunnel facility is utilized to understand the unsteadiness associated with hypersonic axisymmetric cavity flow, owing to its relatively longer test time compared to other impulse facilities. Hypersonic Ludwieg Tunnel (HLT) \citep{karthick2023} located at Technion Israel Institute of Technology, Israel, (see Figure \ref{fig:facility}a) is used for the present set of experiments where the driver tube pressure is varied between $3$ and $10$ bar to generate a wide range of Reynolds numbers. The `U' tube configured driver tube, with an overall length of $3$ m, is connected to a fast-acting valve that separates the tube from the convergent-divergent (C-D) nozzle, designed for a Mach number of $ M_\infty = 6.0$, and the free-jet enclosed-test section. The exit diameter of the nozzle is $D_e=76.2$ mm, out of which the flow is uniform within $60\%$ of the jet core for about $[x/D_e] \approx 1$ \citep{karthick2023} from the nozzle exit. Dry air is used as the test gas for all test cases, except for specific experiments where the test flow is seeded with $CO_2$ to facilitate necessary visualizations. More details about the experimental facility, calibration, and detailed instrumentation can be found in the authors' previous work \citep{karthick2023}. The steady test time for which the flow is expected to be constant can be inferred from the stagnation pressure ($p_{01}$) signal obtained at a location upstream of the C-D nozzle, as shown in Figure \ref{fig:storage press}-I. Observed $[p/p_{01}]$ is constant for $\approx 13$ ms, during which the overall fluctuation level is also limited to $\sigma \approx 0.66\%$. Repeatability of the acquired responses across the runs can also be confirmed from Figure \ref{fig:storage press}-I. 

Local pressure measurements are also performed on an axisymmetric cone without and with a cavity (in terms of a slot) at locations $2.6D$ and $10.6D$ along the cone surface from the leading edge, where $D$ corresponds to the cavity depth of $4$ mm (see Figure \ref{fig:model}). The selected locations are intentional, as they help identify the state of the upstream and downstream boundary layer to the cavity. The fluctuation level is limited to $2\%$ for both the locations on the plain cone (see Figure \ref{fig:storage press} II-a,b), and also the similarity in both the signals is evident, which demarcates the boundary layer to be laminar. With the introduction of the cavity in the cone, the upstream pressure (Figure \ref{fig:storage press} II-c) still pertains to the rms level of $2\%$, whereas the downstream location (Figure \ref{fig:storage press} II-d) indicates a significant fluctuation intensity of about $5.8\%$.

\subsection{Model Details} \label{ssec:model_details}
\begin{figure*}
\includegraphics[width=0.8\textwidth]{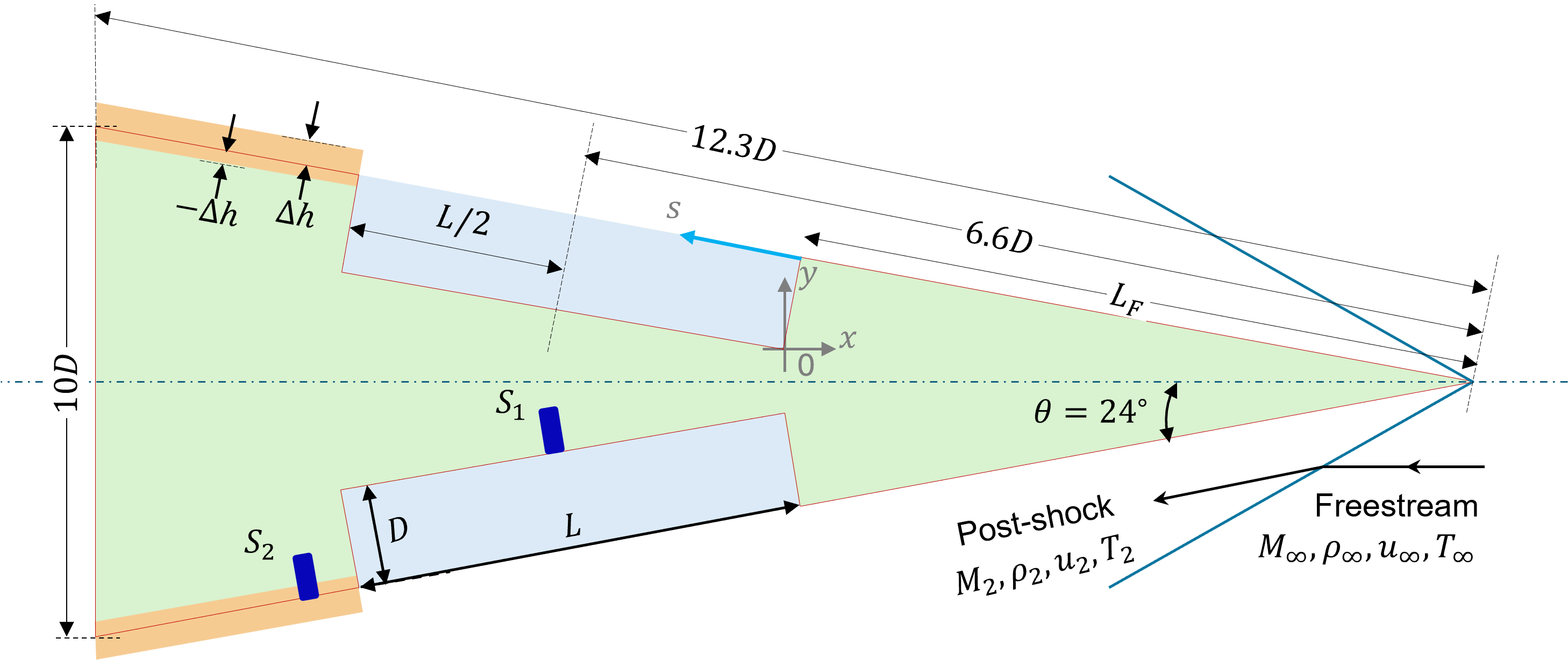}
\caption{Schematic of the cone-mounted cavity configuration, highlighting the key dimensions and cavity nomenclatures used in the present study. $L_F$ will be different based on the $[L/D]$. $S_1$, and $S_2$, correspond to locations of the unsteady pressure probes.}
\label{fig:model}
\end{figure*}

Axisymmetric cavity configurations with desired geometric parameters are carved out in the form of an annular region, having a depth of $D=4$ mm, from a conical model with an apex angle of $2\theta = 48^\circ$ and a base diameter of $40$ mm (see Figure \ref{fig:model}). The slant length of the cavity-mounted cone is $\approx 49$ mm, and the cavity is placed in such a way that the cavity center is always at a fixed distance of $6.6\ D$ from the leading edge, as depicted in Figure \ref{fig:model}. The length-to-depth aspect ratios of the cavity are $[L/D] = [2, 4, 6]$, which correspond to different values of $[L/L_F] =[0.36, 0.87, 1.67]$. In order to assess the effect of the reattachment surface on the overall flow evolution, the height of the trailing edge is decreased as well as increased relative to the leading edge cavity depth ($[\Delta h/D]$) for $[L/D]=6$, which results in $[\Delta h/D] = [-0.5,-0.25,0,0.25,0.5]$. However, in the present study, we mostly concentrate on the cases of $[\Delta h/D] = [-0.25,0,0.5]$, are reported. Experiments are carried out for a set of Reynolds numbers based on the cavity depth $Re_D=[23,\ 30,\ 38,\ 45.2,\ 51.3,\ 59,\ 66.3,\ 74]\times 10^3$ by varying the pressure in the driver tube. All experiments are conducted under room temperature conditions, limiting the total enthalpy to $0.3$ MJ/kg. The flow conditions prevailing in the freestream flow, as well as the cavity upstream, are estimated using isentropic and conical shock relations, which are presented in Table \ref{tab:freestream_cond}.

\begin{table*}
\caption{The derived flow parameters in the freestream and cavity upstream for the extreme cases of driver tube and stagnation pressures ($p_0$ and $p_{01}$), using the isentropic relations.}
\label{tab:freestream_cond}
\begin{ruledtabular}
\begin{tabular}{lll}
\textbf{Parameters} & \textbf{Freestream (${\infty}$)} & \textbf{Cavity upstream ($2$)}\\
\midrule
Static pressure ($p$, Pa) & 152.64 - 514.92 & 1472 - 4785\\
Static temperature ($T$, K) & 36.58 & 91.6\\
Static density ($\rho$, kg/m$^3$) & 0.015 - 0.049 & 0.056 - 0.182\\
Kinematic viscosity ($\nu \times 10^{-5}$, m$^2$/s) & 14.7 - 4.4 & 11.2 - 3.4\\
Velocity ($u$, m/s) & 727.46 & 636.93\\
Shock angle ($\beta$ for $\theta = 24^\circ$, '$^\circ$') & - & 28.3\\
Mach number ($M$, - ) & 6.0 & 3.32\\
Reynolds number ($Re_{D} \times 10^3$, - ) & 19 - 67 & 23 - 74\\
\end{tabular}
\end{ruledtabular}
\end{table*}

\subsection{Measurement Methodology} \label{ssec:meas_meth}

\begin{figure*}
	\includegraphics[width=0.95\textwidth]{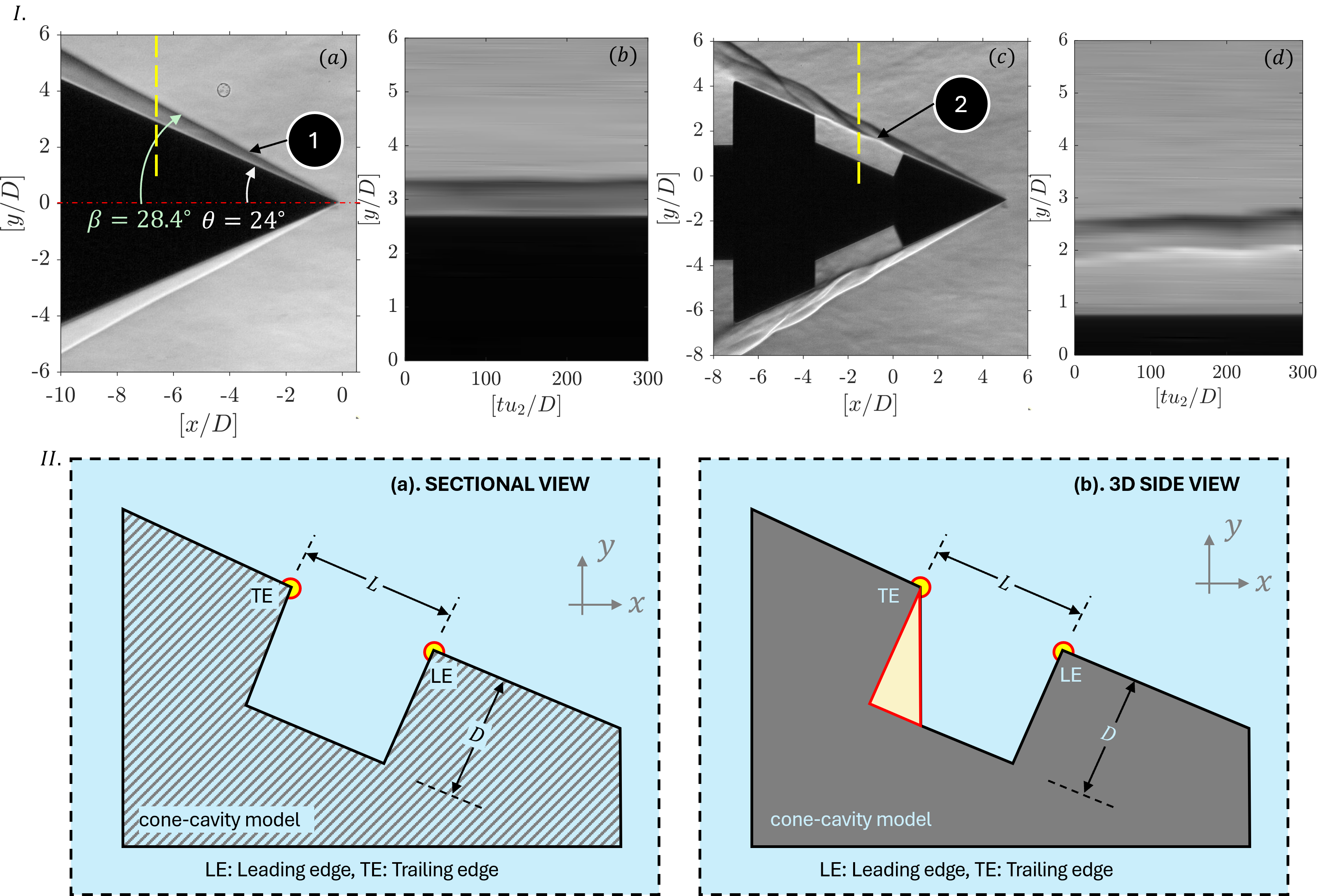}
	\caption{\label{fig:2d_3d_diff} I-(a,c) High-resolution schlieren image obtained for plain cone flow, and cavity mounted cone, respectively with vertical lines marking the profiles along which the space-time plots are construed; I-(b,d) Typical space-time ($y-t$) diagrams, representation of the temporal fluctuations for the respective cases of plain cone and cone-cavity configuration. II-(a) Sectional and (b) 3D side view of the hypersonic cone-cavity to highlight the shadowing of the cavity near the trailing edge while rotating the model about the cone-axis. Cavity configuration corresponds to $[L/D] = 4,\ [\Delta h/D]=0$, and incoming flow $Re_D$ is $74000$. Flow features: 1 - leading-edge shock, 2 - separated shear layer.}
\end{figure*}

The present study involves both high-resolution and high-repetition-rate imaging to identify the overall flow evolution and quantify the associated unsteadiness. High-resolution schlieren imaging (see Figure \ref{fig:facility}b for the set-up) is performed to capture the transverse density gradient ($\partial \rho/\partial y$) through a conventional 'Z-type' setup \citep{settles2001} using an LED light source. A high-speed camera is used to capture the spatial field at a frame rate of $2.2$ kHz corresponding to a resolution of $1280 \times 800$ pixels and an exposure time of $2 \mu$s. Typical schlieren snippet for a plane cone flow and cavity-mounted cone is displayed in Figure \ref{fig:2d_3d_diff} I-a,c. For the semi-cone apex angle of $\theta=24^\circ$, the leading edge shock angle is measured to be $28.4^\circ$ for the flow over plain cone (Figure \ref{fig:2d_3d_diff} I-a), which is a $0.2 \%$ deviation from the estimation using Taylor Maccoll relation. The high-resolution image of the axisymmetric cavity (Figure \ref{fig:2d_3d_diff} I-c) shows undulations along the shear layer, which will be discussed in the subsequent sections. To better understand the unsteadiness, space-time ($y-t$) plots are constructed; however, they may not provide a clear picture (see Figure \ref{fig:2d_3d_diff} I-d) when applied to high-resolution spatial images. The $y-t$ plots (Figure \ref{fig:2d_3d_diff} I-b,d) are generated by stacking the light intensity variation over a desired time interval along the profiles marked by the yellow dashed lines in the Figure \ref{fig:2d_3d_diff} I-a,c, which also corresponds to the location for the cavity center. The leading-edge shock emanating from the cone tip appears fairly steady (Figure \ref{fig:2d_3d_diff} I-b), indicating minimal incoming freestream turbulence to the cavity. 

From the visualization perspective, the axisymmetric cavity geometry seems different (see Figure \ref{fig:2d_3d_diff} II-b) as compared to the sectional view (see Figure \ref{fig:2d_3d_diff} II-a). This is attributed to the rotation of the sectional view about the $x$-axis and the light source being placed sideways to the cone-cavity model, for which the integrated effect looks like the cavity trailing edge being a vertical one as seen in Figure \ref{fig:2d_3d_diff} II-b. Hence, the cavity trailing surface appears to be vertical as shown in Figure \ref{fig:2d_3d_diff} II-b for all the images presented in the subsequent sections, both from schlieren and Rayleigh scattering.

\begin{figure*}
	\includegraphics[width=0.75\textwidth]{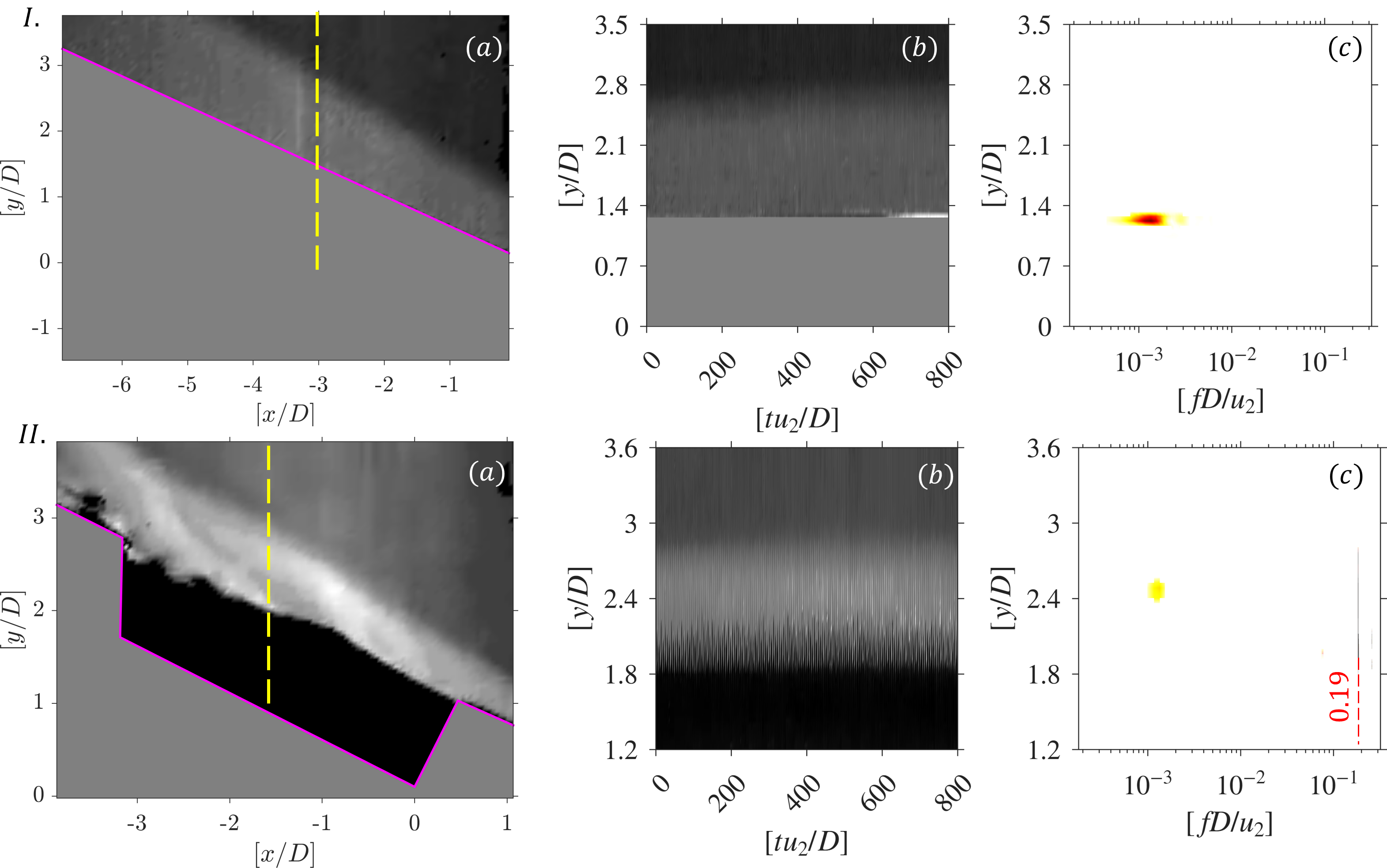}
	\caption{\label{fig:PLRS_xt} Instantaneous (a) PLRS snapshot obtained during high-repetition rate imaging for (I) plain cone flow, and (II) cavity-mounted cone flow. (b, c) represent respective space time ($y-t$), and space-frequency ($y-f$) plots. Cavity configuration corresponds to $[L/D] = 4,\ [\Delta h/D]=0$, and incoming flow $Re_D$ is $74000$. }
\end{figure*}

In contrary to schlieren imaging which yield a line-of-sight-integrated flowfield, planar measurements in a streamwise plane at $z/D=0$ are obtained using the Planar Laser Rayleigh Scattering (PLRS) technique \citep{zhang2016} (see Figure \ref{fig:facility}c for the set-up). \text{$CO_2$} gas is mixed with air in a proportion of $1:9$ by volume in the driver tube, which upon expansion to the freestream conditions forms icicles typically of the order of $0.5\ \mu$m owing to the very low static temperature mentioned in Table \ref{tab:freestream_cond}. PLRS relies on measuring the scattered light intensities from icicles when illuminated by a continuous laser of similar wavelength as the icicle size. It should be noted that it is not possible to capture the low-velocity region within the cavity using PLRS, as icicles tend to evaporate upon interacting with the relatively warmer fluid. High-repetition-rate PLRS images are captured at a resolution of $128 \times 96$, with a fixed acquisition rate of 100 kHz. Typical instantaneous PLRS for the attached flow over the plane cone, and the small-structure laded shear layer can be seen from Figure \ref{fig:PLRS_xt} I-a and \ref{fig:PLRS_xt} II-a, respectively. In a similar way as mentioned previously, $y-t$ plots (Figure \ref{fig:PLRS_xt}b) are generated along with the corresponding $y-f$ plot (Figure \ref{fig:PLRS_xt}c) through fast Fourier transform, depicting the dominant spectral content. No susceptible frequency is identified for the plane cone flow, as the broadened spectra seen in Figure \ref{fig:PLRS_xt} I-c is attributed to an anomaly of laser light reflection on the cone surface. However, a dominant frequency is identified in the case of an axisymmetric cavity (Figure \ref{fig:PLRS_xt} II-c), thereby indicating the capability to resolve the unsteadiness across the shear layer. More details about the schlieren and PLRS system can be found in Karthick et al. (2023) \citep{karthick2023}. 

Local mean pressure variation and pertinent dominant spectral content across the cases are quantified using the unsteady pressure sensors (Kulite\textsuperscript{\textregistered} XCQ-080-100A) mounted on the cavity floor center ($S_1$ as depicted in Figure \ref{fig:model}) and on the cone surface downstream of the flow reattachment ($S_2$ which is at $10.5D$ from the leading edge). Pressure responses are recorded using NI data-acquisition cards at a sampling rate of $800$ kHz.

\subsection{Uncertainty Quantification}

Uncertainty estimation is a critical aspect to ascertain the variation in the parameters derived during the experimental campaign. The analysis accounts for contributions to uncertainty from systematic and random sources, including measurement accuracy, run-to-run variability, background noise, geometric tolerances, spatial resolution, and statistical dispersion associated with visually identified features \citep{Coleman2009}. The overall uncertainty in any parameter is calculated by aggregating individual contributing sources using a statistical-based root-sum-square methodology \citep{golui2025}. Primarily, the uncertainty in freestream conditions such as Mach number, pressure, and temperature is calculated based on Rayleigh-pitot and isentropic relations in conjunction with uncertainty in the driver tube pressure ($p_{01}$) and pitot pressure ($p_{02}$) measurement. Effectively, the uncertainty in freestream Reynolds number is calculated to be $\pm 3.4\%$. The high-repetition PLRS snapshots at the current settings yield a spatial resolution of $0.22$ mm, resulting in an uncertainty of $\pm 4.8\%$ in the estimated onset location of convective structures in the shear layer. The frequency resolution of the PLRS-based estimate is $167$ Hz, calculated from $600$ frames during the test duration, whereas the resolution of the sensor-based estimate is $83$ Hz. As a result, the variation in $St_D$ is expected to be $\pm 3.5\%$ and $\pm 2.5\%$ for PLRS and sensor-based estimate, respectively. The uncertainty in non-dimensional mean pressure variation, which is mainly contributed by the standard deviation across the runs, is about $\pm 3.1\%$. 

\section{Results and Discussion}

\begin{figure*}
	\includegraphics[width=0.75\textwidth]{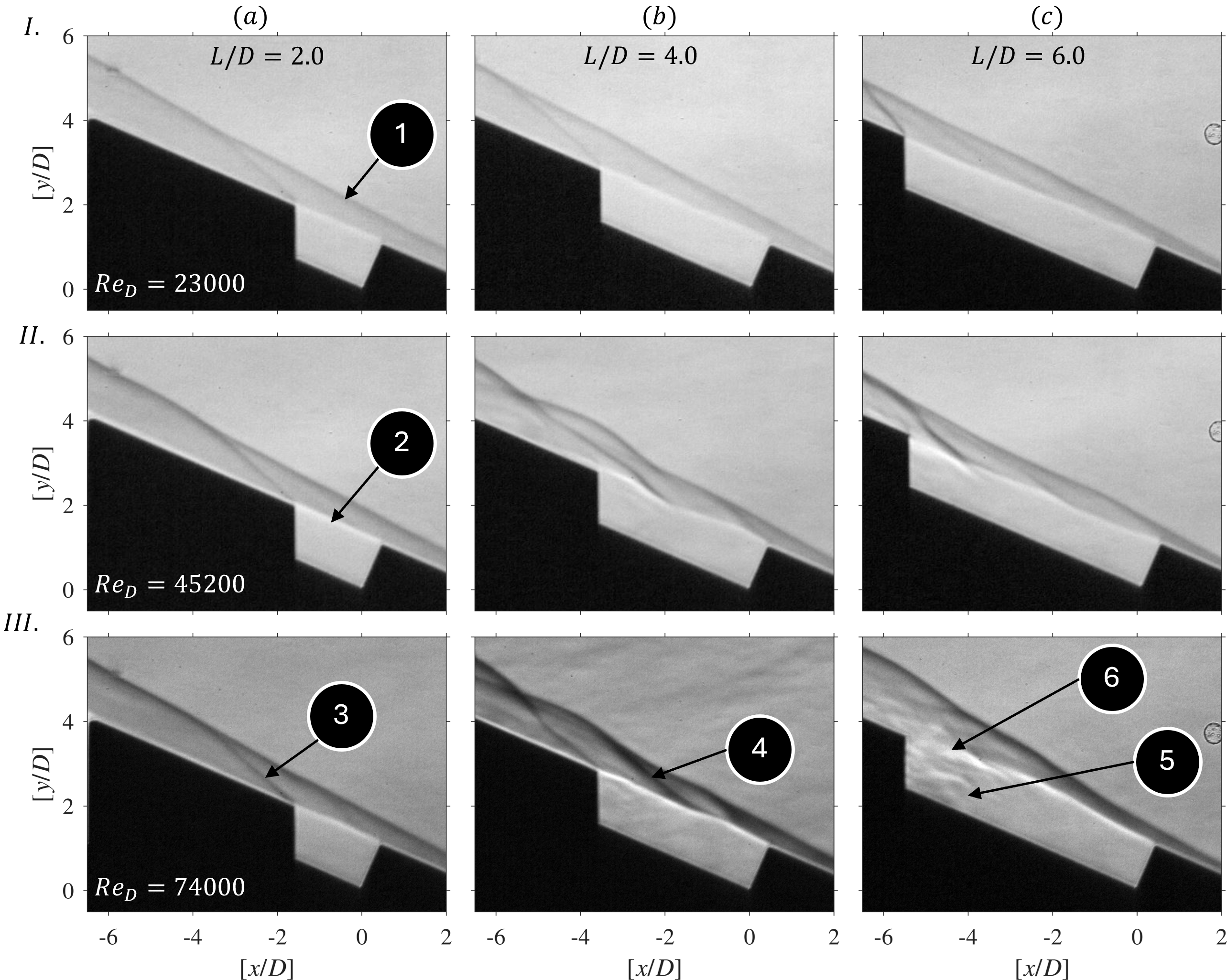}
	\caption{\label{fig:sch_ld} Effect of length-to-depth ratio ($[L/D]$) at different Reynolds number ($Re_D$) on the overall flow features obtained through high-resolution schlieren imaging. Instantaneous schlieren images taken at an arbitrary time-step for different cases with no-excess trailing edge height case ($[\Delta h/D]=0$) for: (a) $[L/D]=2.0$, (b) $[L/D]=4.0$, and (c) $[L/D]=6.0$. Each row corresponds to a fixed value of Reynolds number where (I) $Re_D = 23000$, (II) $Re_D = 45200$, and (III) $Re_D = 74000$. Key flow features: 1. leading edge conical shock, 2. separated shear layer, 3. reattachment shock at the trailing edge, 4. shock wave from the undulating shear layer, 5. upstream propagation of acoustic waves, and 6. shear layer turbulent break-down. See the \href{https://youtu.be/IWpR-vujNLw}{supplementary} for viewing the corresponding video files (top row is PLRS, bottom row is schlieren).}
\end{figure*}

Prior to experiments with the axisymmetric cavity, it is intended to gather information on the prevailing flow state on the cone surface. The high-resolution schlieren snapshot, along with an instantaneous PLRS image for hypersonic flow over the cone, is depicted in Figure \ref{fig:2d_3d_diff} I-a and Figure \ref{fig:PLRS_xt} I-a. It is clear from both figures that the flow remains attached to the entire cone surface and is free of any structures, thereby indicating that the flow is laminar. Also, the $y-f$ analysis (Figure \ref{fig:PLRS_xt} I-c) portrays no significant spectral content. From Figure \ref{fig:storage press} II-a,b, it can also be ensured that the pressure values downstream of the leading edge shock have good agreement with those of the estimated conditions mentioned in Table \ref{tab:freestream_cond}, and the fluctuation levels are within $\pm 2\%$. Therefore, irrespective of the cavity location on the cone surface, laminar flow is expected at the cavity entrance. In the following section, the effects of geometric parameters and incoming flow conditions on the flow evolution are discussed, utilizing both qualitative and quantitative perspectives.

\subsection{Effect of aspect ratios, $[L/D]$} \label{ssec:aspect_ratio}

\subsubsection{Qualitative interpretation}

\begin{figure*}
	\includegraphics[width=0.85\textwidth]{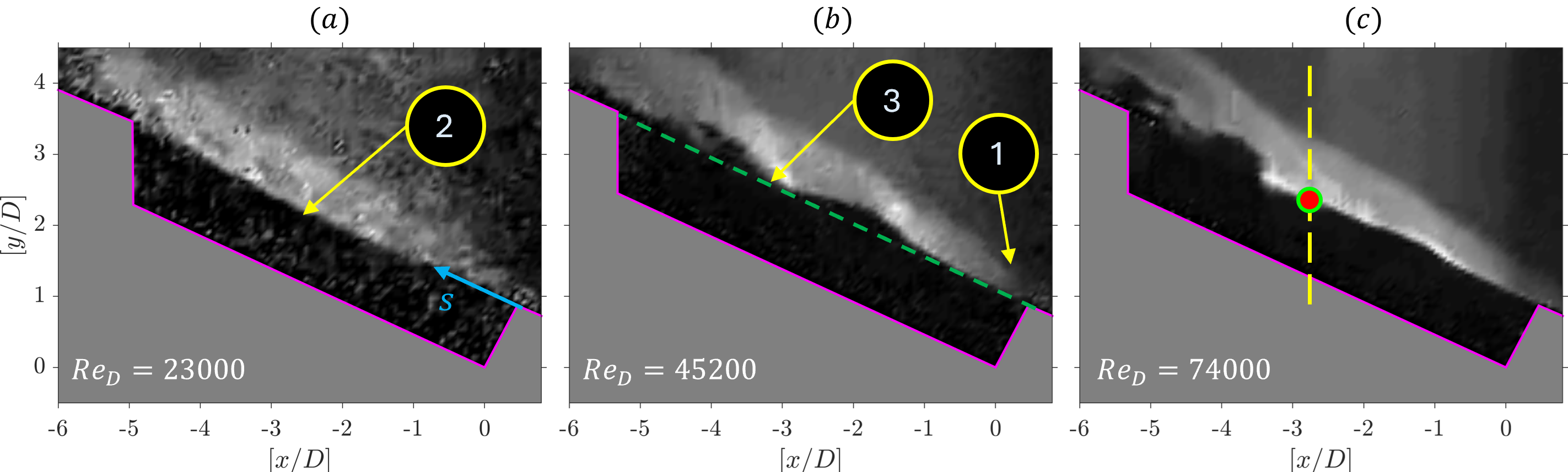}
	\caption{\label{fig:PLRS_typical} Typical PLRS snapshot for different $Re_D$s of (a) $23000$, (b) $45200$, and (b) $74000$ obtained for the case of $[L/D]=6$. Key flow features: 1. conical shock from the leading edge, 2. laminar shear layer, 3. K-H vortices. See the \href{https://youtu.be/VRJ5KJEjSIE}{supplementary} for viewing the corresponding video files. }
\end{figure*}

The schlieren images obtained using high spatial resolution imaging for several $[L/D]$ and $Re_D$, having $[\Delta h/D]=0$, are shown in Figure \ref{fig:sch_ld}. Instantaneous images are used to identify the gross flow features, allowing for the qualitative drawing of conclusions on the intensity of shock-related unsteadiness. Some of the key flow features evident from the schlieren snapshots are: the leading edge shock from the cone tip (marked as \circled{\small 1} in Figure \ref{fig:sch_ld}), which makes the flow parallel to the cone surface, flow separation leading to the formation of a separated shear layer at the front cavity face (marked as \circled{\small 2} in Figure \ref{fig:sch_ld}), and the reattachment shock emanating from the trailing edge (marked as \circled{\small 3} in Figure \ref{fig:sch_ld}). All $[L/D]$ cases considered herein across the $ Re_D$ range portray a separated shear layer spanning the entire cavity length, indicating the typical characteristics of an open cavity configuration. Irrespective of the $Re_D$, a laminar separated shear layer with no evidence of vortex roll-up is perceived for $[L/D]=2$ (Figure \ref{fig:sch_ld}a), which is also true for the lowest $Re_D = 23000$, across the $[L/D]$ cases (see Figure \ref{fig:sch_ld}-I). For $[L/D] = [4,6]$, further increase in $Re_D$ results in the onset of convecting vortical structures (Figure \ref{fig:sch_ld} II-b) along the laminar shear layer, caused by Kelvin-Helmholtz (K-H) instability. These vortical structures is accompanied by weak Mach waves along the convecting direction of the supersonic stream (marked as \circled{\small 4}), as well as acoustic waves in the recirculating subsonic zone. Also, the shock and separated shear layer motion intensify with an increase in $Re_D$, thereby significantly enhancing the reattachment shock's strength (Figure \ref{fig:sch_ld} II-c, Figure \ref{fig:sch_ld} III-b). At the highest $Re_D = 74000$ for $[L/D]=6$, acoustic waves originating from the trailing edge (marked as \circled{\small 5} in Figure \ref{fig:sch_ld}) and propagating into the cavity upstream are clearly visible (See the \href{https://youtu.be/IWpR-vujNLw}{supplementary} for the schlieren video acquired in high-repetition rate). This, in turn, amplifies the K-H vortices on the separated shear layer to grow spatially along the streamwise direction and eventually break down to a turbulent state. Disappearance of waviness in the separated shear layer marks the ending of the laminar shear layer (compare the white stripe in Figure \ref{fig:sch_ld} III-b,c, also marked as \circled{\small 6} in Figure \ref{fig:sch_ld}). As only three cases of $Re_D$ are shown in Figure \ref{fig:sch_ld} for brevity, the exact transitional $Re_D$ will be estimated during the analysis of high-speed PLRS snapshots. As mentioned earlier in Eq. \ref{eq:unsteadiness_identify},  Creighton \& Hillier \citep{creighton2007} developed a criterion to determine the critical ratio of $[L/L_F]$ beyond which unsteadiness in the shear layer can be expected. Based on Eq. \ref{eq:unsteadiness_identify}, for the current experimental scenario, the estimated critical $[L/L_F]$ is $0.568$. In agreement with their prediction, the case of $[L/D]=2$, for which the $[L/L_F]=0.36$ indeed shows a steady flowfield, whereas for $[L/D]=[4,6]$, the corresponding $[L/L_F]=[0.87, 1.67]$ are greater than the critical $[L/L_F]$ and both shear layers show unsteadiness.

\begin{figure}
	\includegraphics[width=0.9\columnwidth]{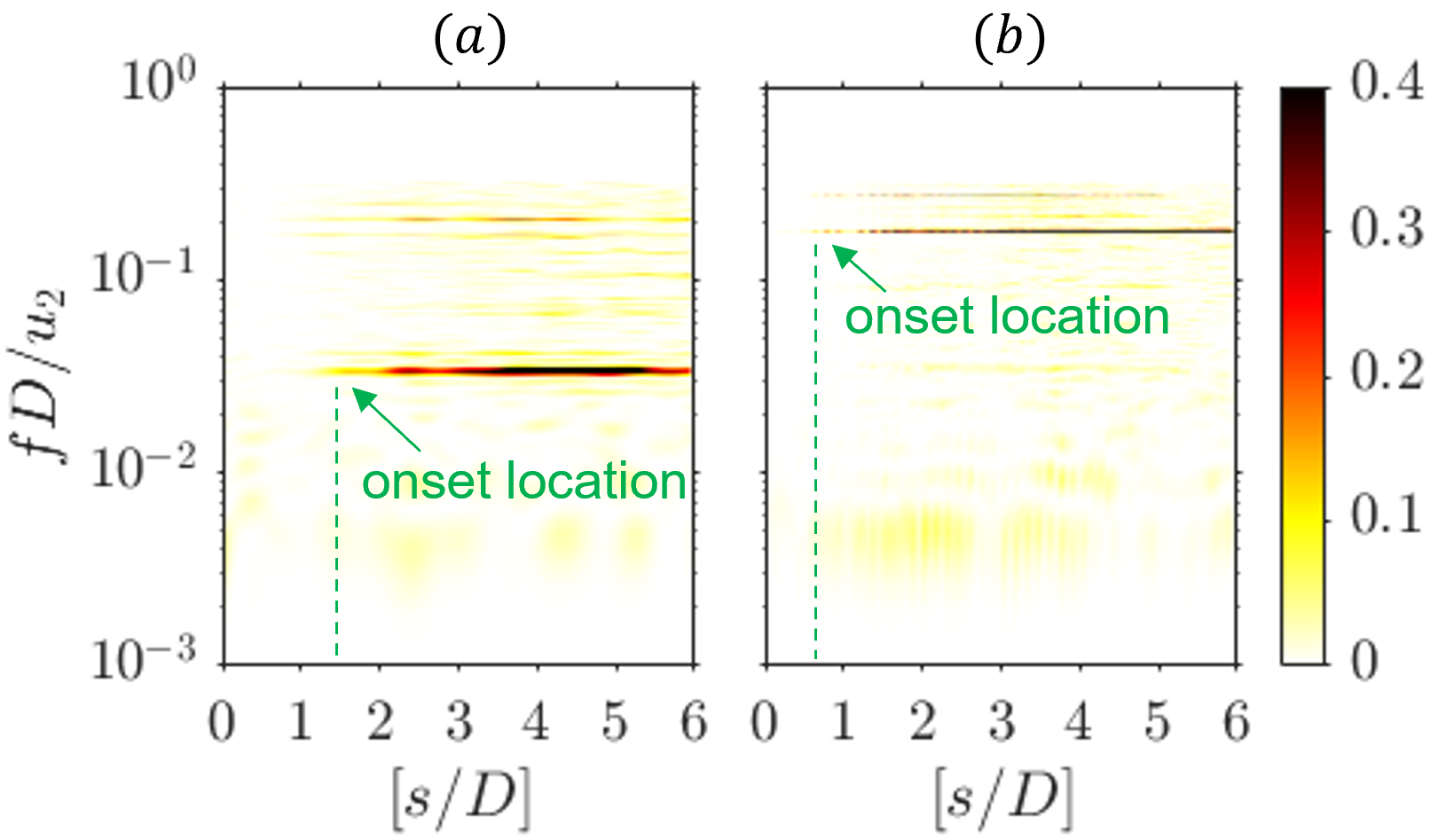}
	\caption{$s-f$ diagram generated for the case of $[L/D] = 6$ having (a) $Re_D = 23000$, and (b) $Re_D = 74000$ where the profile is generated along the shear layer. The respective line is marked in dashed green color in Figure \ref{fig:PLRS_typical}b.}
    \label{fig:xf_LD6}
\end{figure}

Schlieren imaging is a line-of-sight integrated technique that does not capture some flow features, such as three-dimensional small-scale structures or the transition point in the separated shear layer in a given plane. Therefore, PLRS imaging is carried out for all the cases across $Re_D$s but only shown for $[L/D]=6$ in Figure \ref{fig:PLRS_typical} at specific $Re_D$ for brevity. It is noted that, for the lowest $Re_D$ case, the snapshot appears hazy (due to poor light intensity) caused by the low-density field resulting from the low fill pressure conditions in the driver tube. A laminar shear layer (marked as \circled{\small 2} in Figure \ref{fig:PLRS_typical}) spanning the cavity (Figure \ref{fig:PLRS_typical}a) is seen for most of the cases. However, for an increased $Re_D$ for $[L/D]=4$, small-scale  K-H vortices are seen to be convected along the separated shear layer (see Figure \ref {fig:PLRS_xt} II-a). For a fixed $Re_D$, the wavelength of the K-H vortices increases as $[L/D]$ increases (compare Figure \ref{fig:PLRS_xt} II-a and Figure \ref{fig:PLRS_typical}c), which is in line with earlier observations\citep{doshi2022}. Moreover, the point of origin of these K-H vortices, where the shear layer is perturbed by the acoustic wave, seems to move further upstream along the shear layer as $[L/D]$ increases (compare Figure \ref{fig:PLRS_xt} II-a, and Figure \ref{fig:PLRS_typical}c). The shear layer perturbed location can be demarcated from the $s-f$ plots as shown in Figure \ref{fig:xf_LD6}, where the line profile considered is along the shear layer as marked in Figure \ref{fig:PLRS_typical}b, , where $s$ is the line along the cavity shear line starting at the leading edge (see Figure \ref{fig:PLRS_typical}a,b). With increases in $Re_D$ from $23000$ to $74000$, the dominant frequency in each of the cases starts appearing from $[s/D]$ of $1.5$ and $0.6$, respectively. These estimates can be considered as the probable location for perturbation, as upstream of it, a laminar shear layer exists. With an increase in $Re_D$, the upstream movement of the perturbation location indicates the availability of a longer length (till the rear face of the cavity) for instability wave amplification and breakdown into turbulence. Nevertheless, with an increase in $Re_D$ for $[L/D]=6$ (see Figure \ref{fig:PLRS_typical}c), the wavelength of the K-H vortices is shortened, and their coherence seems to be fading out at some downstream point, thereby indicating the possibility of shear layer breakdown, which is apparent from the schlieren snapshot and upstream shear layer perturbation.

\subsubsection{Quantitative Estimation}\label{sec:quant L_d var}

\begin{figure*}
	\includegraphics[width=0.7\textwidth]{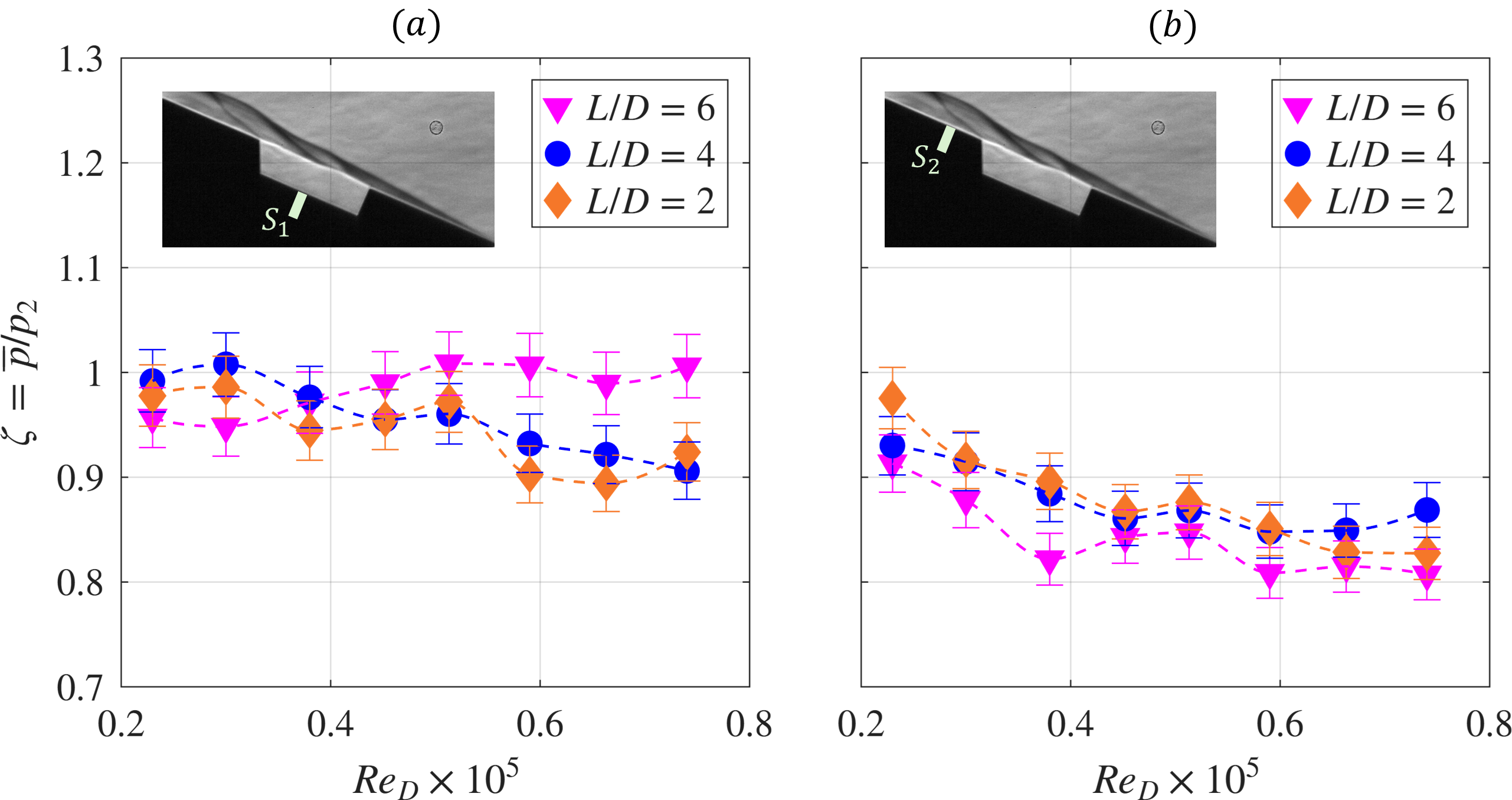}
	\caption{Non-dimensional mean pressure variation at the (a) cavity floor center, $S_1$, and (b) cavity downstream, $S_2$, location with change in $[L/D]$ and $Re_D$.}
    \label{fig:p_LbyD_effect}
\end{figure*}

The effect of cavity aspect ratio ($[L/D]$) with $[\Delta h/D] = 0$ is quantified for a range of $Re_D$ in terms of the mean pressure variation ($\zeta = \overline{p}/p_2$) at the cavity floor ($S_1$) and downstream cavity location ($S_2$) as shown in Figure \ref{fig:p_LbyD_effect}. For all cavity aspect ratios ($[L/D]$) considered herein, the mean pressure magnitude on the cavity floor varies in the range of $0.9\leq \zeta \leq 1.0$ (see Figure \ref{fig:p_LbyD_effect}a), thereby signifying that the recirculation bubble entrapped by the separated shear layer and the cavity floor are almost in pressure equilibrium. Moreover, at lower $Re_D$s, the pressure variation between $[L/D]$ cases is relatively lesser, and at the highest range of $Re_D$s, the percentage of variation in $\zeta$ increases. For $[L/D] = 6$, at higher $Re_D$s, the $\zeta$ is found to be almost 1, which is probably a result of better mixing and interaction with the freestream flow due to the turbulence breakdown of the shear layer as seen during the high-resolution schlieren imaging (Figure \ref{fig:sch_ld} III-c).

\begin{figure*}
	\includegraphics[width=0.7\textwidth]{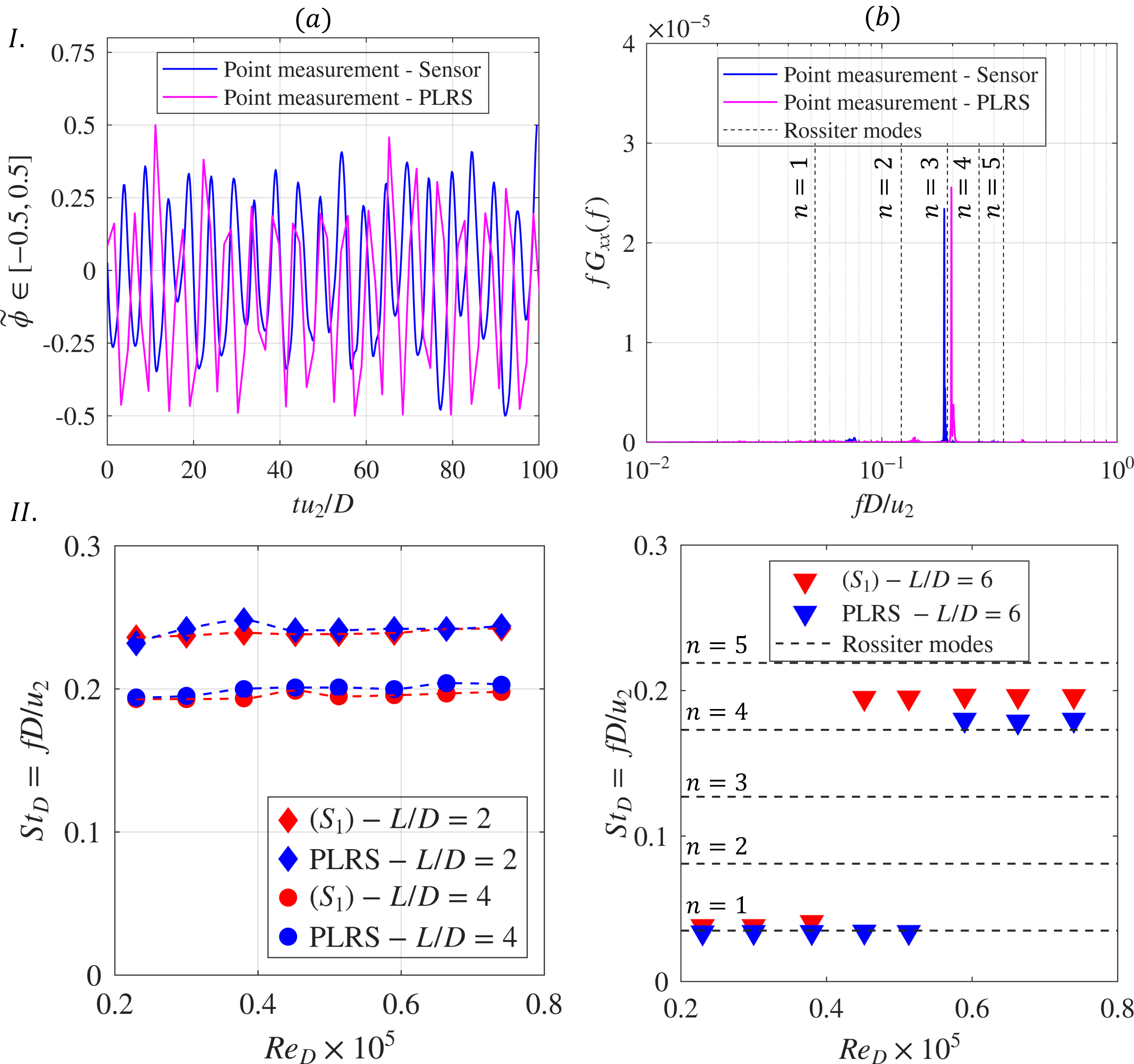}
	\caption{(I-a) Rescaled fluctuation levels ($\tilde{\phi}$) obtained during the run-time from PLRS imaging on the shear layer (see Figure \ref{fig:PLRS_typical}c for the location) and from the sensor $S_1$ placed on the cavity floor center for the case of $[L/D]=4$ at $Re_D=74000$. (I-b) The spectra obtained using FFT on the normalized fluctuations signal depicting the dominant $St_D$ from both the techniques and their agreement with the predicted frequencies of the Rossiter modes. (II) Comparison of dominant $St_D$ obtained from sensor $(S_1)$ and PLRS image based processing for (a) $[L/D] = [2.0, 4.0]$, and (b) $[L/D] = 6$. The dashed black lines represent the Rossiter modes obtained using the semi-empirical relation (Eq. \ref{eq:rossiter}).}
    \label{fig:sensor_PLRS_freq_LD}
\end{figure*}

Similarly, at the downstream cavity location $S_2$, owing to the interaction of the vortices with the trailing edge followed by the reattachment shock oscillation, inability of the pressure recovery is evident for $[L/D] = [2, 4, 6]$ at most $Re_D$s, thereby resulting in a lesser mean pressure downstream of the cavity (Figure \ref{fig:p_LbyD_effect}b). Irrespective of the $[L/D]$, an overall decrement in magnitude for $S_2$ location is observed with increases in $Re_D$. For lower $Re_D$ cases, weaker density gradient near the trailing edge (Figure \ref{fig:sch_ld} I,II-a) indicates possibility of gradual flow turning, and more effective pressure recovery leading to similar magnitudes at the $S_1$ and $S_2$ locations across $[L/D]$s. With increment in $Re_D$, owing to the occurrence of the convecting K-H vortices (Figure \ref{fig:sch_ld} II,III-b), reattachment happens over an extended streamwise region followed by distributed flow turning and expansion, leading to lesser $\zeta$. Moreover, apart from $[L/D] = 6$, all the other cases experience less than $\pm 8\%$ variation between pressure level on the $S_1$ and $S_2$ location across the entire $Re_D$ range considered herein. For $[L/D] = 6$, the $S_2$ location is found to have $\approx 20\%$ lesser mean pressure in comparison to the $S_1$ for higher $Re_D$s. At higher $Re_D$s, the shear layer transitions to turbulence, resulting in a weaker reattachment shock than in other cases, which can explain lower pressure at the $S_2$ location.

\begin{figure*}
	\includegraphics[width=0.85\textwidth]{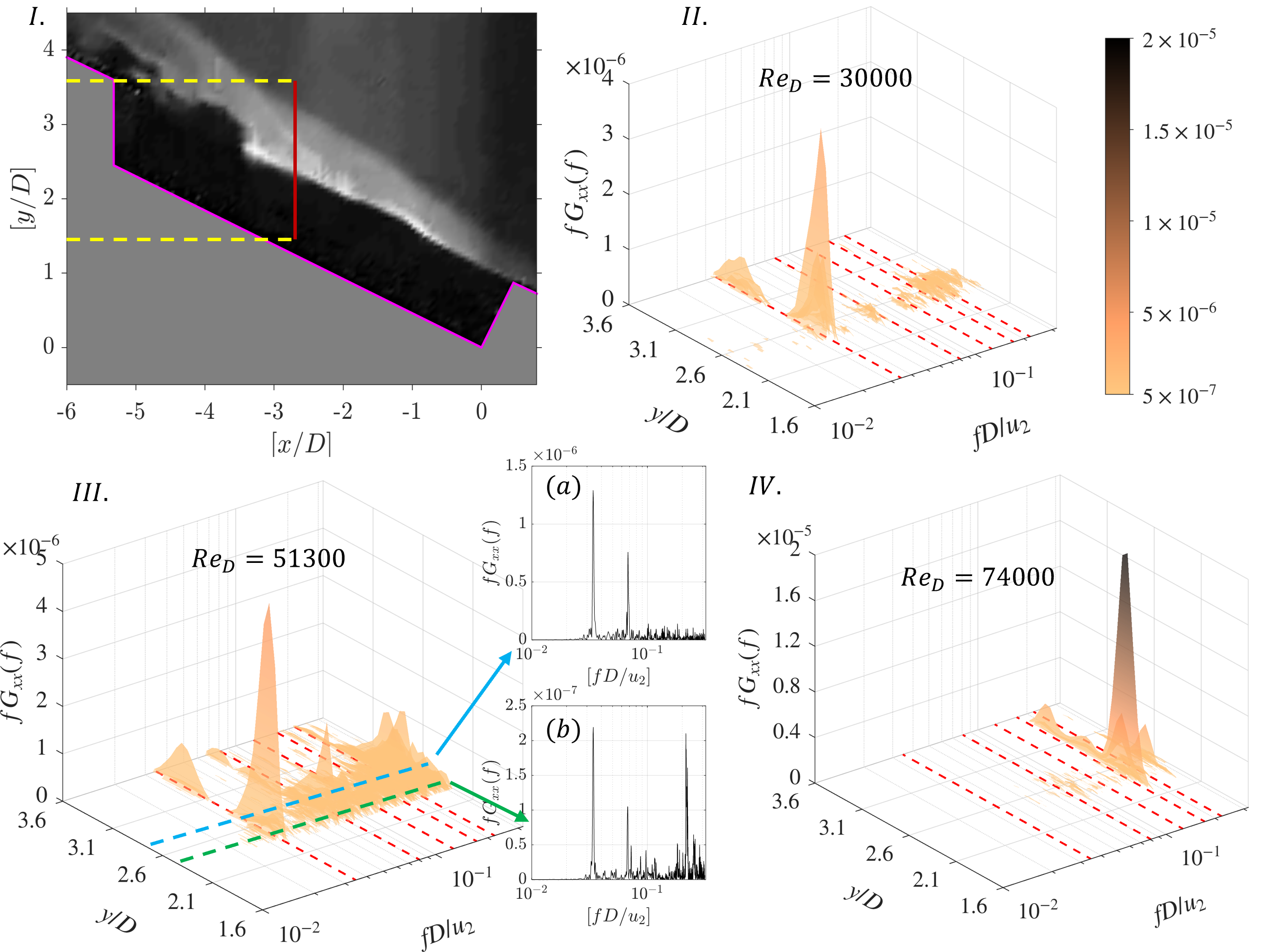}
	\caption{ (I) Representational vertical line in the cavity center marked in red color used for constructing the surface plot depicting the variation of power spectrum at different spatial locations ($[y/D]$) for (II) $Re_D = 30000$, (III) $Re_D = 51300$, and (IV) $Re_D = 74000$. The red dashed lines represent different Rossiter modes. For $Re_D = 51300$, two sub-figures (a, b) are presented which depict the spectra at two locations within the shear layer, confirming the presence of multitude modes.}
    \label{fig:surf_freq_yd}
\end{figure*}

The dominant frequency is identified from pressure responses measured at location $S_1$ during the run-time through FFT analysis. In parallel, to support the analogy from the unsteady visualization, a point on the cavity shear layer at the cavity centerline is considered (see Figure \ref{fig:PLRS_typical}c), and the temporal variation of the light intensity is extracted from the PLRS snapshots and then subjected to the FFT procedure. Before discussing the variation in the dominant frequency, it is essential to establish that the measured unsteadiness is consistent and exhibits a dominant spectrum. The light-intensity fluctuations obtained from PLRS and the pressure fluctuations measured at the cavity center are shown in Figure \ref{fig:sensor_PLRS_freq_LD} I-a, for the case of $[L/D]=4$ at $Re_D=74000$. The associated spectra are also presented in Figure \ref{fig:sensor_PLRS_freq_LD} I-b, which displays a close match between both the techniques as well as with the predicted frequency of the Rossiter mode within the measurement uncertainty. Based on the agreement, the comparative assessment of dominant frequencies obtained using both methods is plotted in Figure \ref{fig:sensor_PLRS_freq_LD} II-a for $[L/D] = [2, 4]$. Irrespective of the $[L/D]$, the dominant frequency is found to be almost invariant of $Re_D$. Also, the frequency values obtained from the sensor-based FFT analysis and PLRS-based FFT analysis show a close agreement (see Figure \ref{fig:sensor_PLRS_freq_LD} II-a). Finally, the identified frequencies for $[L/D] = [2, 4]$ match the corresponding 2nd ($n=2 \to St_D = 0.242$) and 3rd ($n=3 \to St_D =0.19$) Rossiter modes, respectively as calculated using Eq. \ref{eq:rossiter}. It is also to be noted that the Rossiter formula was originally derived for a 2D cavity, and the good agreement with the present axisymmetric cavity is probably due to the relatively thin shear layer compared with the radius to the cone centerline. It is evident from Eq. \ref{eq:rossiter} that the resonant frequency is inversely proportional to the cavity length. As a result for $[L/D] = 2$, the acoustic feedback time will be shorter, resulting in a higher dimensional frequency, which can also be referred from Figure \ref{fig:sensor_PLRS_freq_LD} II-a. In conjunction with the high-resolution snapshots (see Figure \ref{fig:sch_ld}b), it can be deduced that for $[L/D]=4$, the availability of longer streamwise extent enables instability growth (at higher $Re_D$s), and the cavity resonance remains the driving factor, resulting in the $St_D$ invariance of $Re_D$.

For the case of $[L/D] = 6$, a change in the dominant non-dimensional frequency is observed as the $Re_D$ is increased (see Figure \ref{fig:sensor_PLRS_freq_LD} II-b). Furthermore, there is a disparity between PLRS-based analysis and sensor-based FFT estimation in indicating the $Re_D$ value at which mode switching occurs. From the shear layer measurements through PLRS, the associated dominant $St_D$ is seen to be increased from $St_D = 0.035$ to $St_D = 0.19$, and the $Re_D$ corresponding to mode switching is found to be $Re_D\geq 51300$. However, the mode switching is evident for $Re_D\geq 38000$ from the FFT estimated at the cavity floor center ($S_1$). At intermediate $Re_D$s, both the flapping (Rossiter $1$st mode), dominating the low Re range, and the K-H (Rossiter $4$th mode), dominating the high Re range, coexist. Therefore, at this range, the dominating spectral peak might switch owing to the probing location. The shear layer apparently contains fine-scale turbulent-like structures at higher $Re_D$s (see Figure \ref{fig:sch_ld} III-c and Figure \ref{fig:PLRS_typical}c), in comparison to the large scale K-H vortices embedded in the laminar shear layer at intermediate $Re_D$s (see Figure \ref{fig:sch_ld} II-c and Figure \ref{fig:PLRS_typical}b). It is also worth noting that the $St_D$ inferred from the PLRS imaging agrees closely with the Rossiter modes better than the pressure-based estimate. This might be due to the several reasons among which the azimuthal structure of the disturbances can be a factor which will be described later during the discussion on the $\Delta h/D$ variation.

\begin{figure*}
	\includegraphics[width=0.9\textwidth]{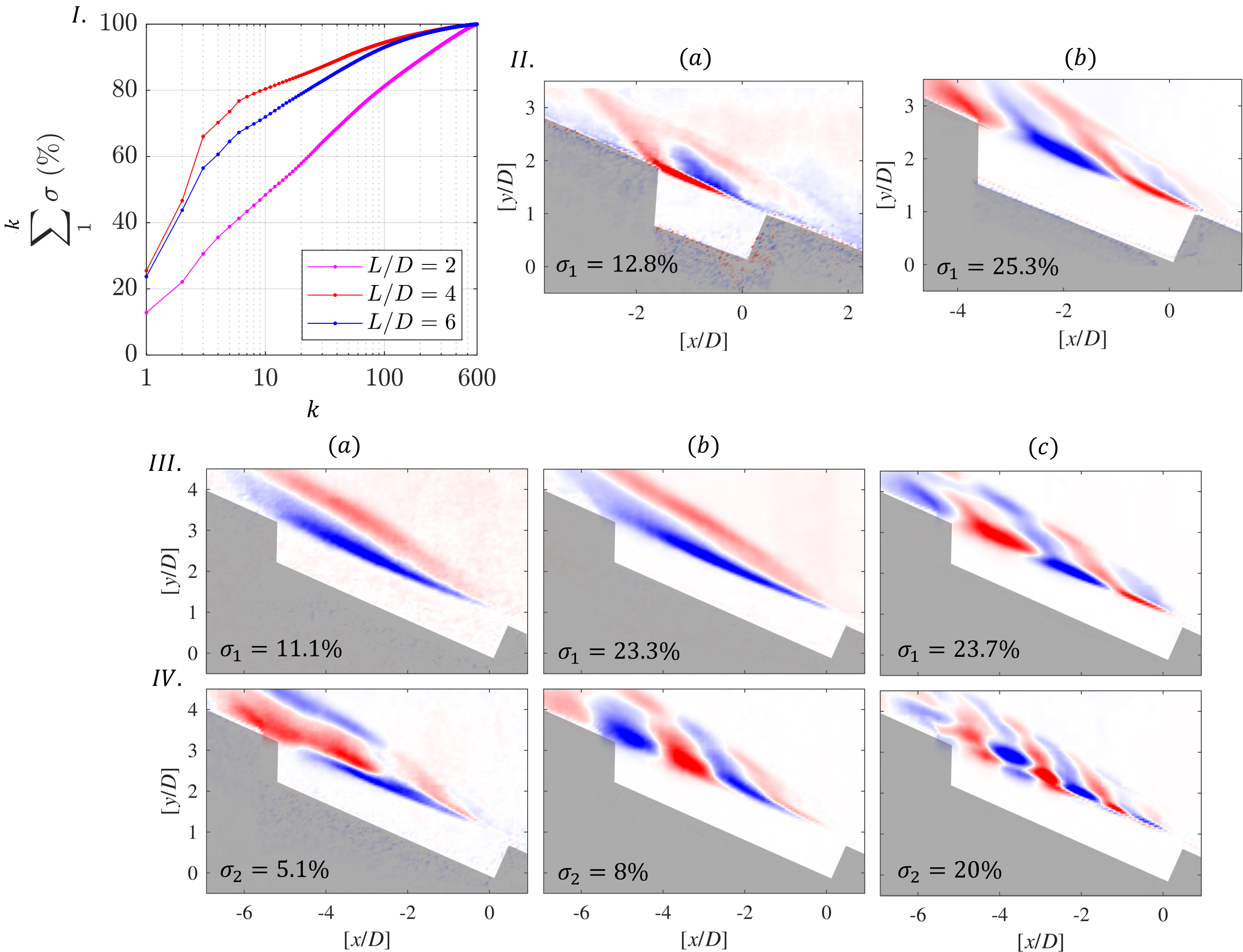}
	\caption{(I) Variation of cumulative energy content obtained from POD analysis for different $[L/D]$ at $Re_D=74000$. $k$ represents the number of instantaneous images that are used for getting the dominant modes, which is 600 for the present case. $\sigma$ is representational of the associated energy content arranged in descending order. (II) Dominant spatial modes obtained through POD for (a) $[L/D]=2$, (b) $[L/D]=4$. $\sigma_1$ represents the energy content of the dominant mode for the respective $[L/D]$. Leading (III) first, and (IV) second POD modes representational of the overall flow for $[L/D]=6$ at different $Re_D$ of (a) $30000$, (b) $51300$, and (c) $74000$. Modal energy fractions $\sigma_{1}$ and $\sigma_{2}$ are indicated in each of the spatial modes.}
    \label{fig:pod_Re_D_ld}
\end{figure*}

In order to ascertain the presence of multiple modes at intermediate $Re_D$s which is the attributing factor for the disagreement between the imaging-based and sensor-based analysis, a surface plot (see Figure \ref{fig:surf_freq_yd}) is prepared by extracting the power spectrum at all the spatial locations along a vertical line passing through the cavity center (see Figure \ref{fig:surf_freq_yd}-I). At $Re_D = 30000$ (Figure \ref{fig:surf_freq_yd}-II), the spectrum at all $[y/D]$ locations within the shear layer thickness exhibits a dominant peak corresponding to the $1$st Rossiter mode, which is in line with earlier observations. In a similar context, for $Re_D = 74000$ (Figure \ref{fig:surf_freq_yd}-IV), the high spectrum amplitude is clearly aligned with the $4$th Rossiter mode, indicating the sole dominance of fine-scale structures. However, at $Re_D = 51300$ (Figure \ref{fig:surf_freq_yd}-III), the surface plot displays a mixture of several spectral peaks with the dominant one pertaining to $n=1$ of Rossiter modes. In order to ensure the presence of wave structures pertaining to different Rossiter modes, spectra obtained at two vertical locations within the shear layer are shown in Figure \ref{fig:surf_freq_yd} III-a,b. In these  range of intermediate range of $Re_D$s, both modes (Rossiter $1$st and $4$th mode) co-exist. The PLRS is sensitive to the exact location along and across the shear layer, whereas the pressure sensor is not. In both cases we see the switching of modes, but at a different $Re_D$. Figure \ref{fig:surf_freq_yd} III shows that in these intermediate range of $Re_D$s we have multiple modes, unlike the lower and higher $Re_D$ cases, and that the peak spectra in the PLRS depend on the exact position of the measurement. 

\subsubsection{Dominant Spatial Structure Estimation}

Discrete spectra observed across different $[L/D]$ and $Re_D$ combinations, as well as their association with various Rossiter modes, necessitate estimating the dominant spatial feature in each case. To estimate the spatial structures, Proper Orthogonal Decomposition (POD) \cite{lumley1967,sirovich1987,Meyer2007} is used in the current study. The instantaneous high-speed PLRS images acquired during the test time are used for this analysis, where each snapshot is arranged as a column of a data matrix ($X$), from which the temporal mean is subtracted to obtain the fluctuation matrix ($X_1$). The covariance matrix ($C = X_1^T X_1$) was then formed and its eigenvalues ($\lambda_i$) and eigenvectors ($v_i$) computed via eigen decomposition ($C v_i = \lambda_i v_i$). The eigenvalues sorted in descending order were then used to determine the cumulative energy distribution across modes, while the spatial modes ($\psi_i$) were obtained by projecting the eigenvectors onto the fluctuation data and normalizing the same ($\psi_i = {X_1 v_i}/{\|X_1 v_i\|}$).

The cumulative energy trend (Figure \ref{fig:pod_Re_D_ld}-I) seems to be linearly increasing for the $[L/D]=2$, whereas quasi-exponentially for $[L/D]=[4,6]$, thereby indicating a significant contribution from the primary modes. It is also clear that the associated energy level for the first dominant mode is $\approx25\%$ for higher $[L/D]$s, whereas $13\%$ for $[L/D]=2$. For $[L/D] = 2$, as the separated shear layer is mostly laminar, with very low amplitude K-H vortices embedded within it, which led to an almost linear increment in the energy content, with the requirement of a significantly higher number of modes to represent $60\%$ of the total energy content. However, to represent $60\%$ of the total flow energy, only $3$ modes are required for $[L/D]=4$, whereas $4$ modes are essential for $[L/D]=6$. Due to the shear layer breakdown to turbulence at $[L/D, Re_D] = [6,74000]$, the fine-scale structures might have led to a lower slope for the increase in the cumulative energy. 

The dominant spatial mode obtained through eigenvectors projection for different $[L/D]$ is presented in Figure \ref{fig:pod_Re_D_ld}-II and \ref{fig:pod_Re_D_ld} III-c. It is evident that irrespective of $[L/D]$ at $Re_D = 74000$, the dominant mode consists of K-H vortices convecting along the shear layer, where alternate red and blue patches represent the streamwise wavelength of the convective structures. It is also clear that for $[L/D] = [2,4,6]$ the number of K-H vortices required to cover the entire length of the cavity is estimated to be $2, 3$, and $ 4$, respectively. The number of structures is also representative of the Rossiter mode expected to dominate the flow. This is in line with the earlier discussion (see Sec. \ref{sec:quant L_d var}) that for $[L/D] = [2,4]$, the dominant discrete frequency in the spectra is closer to the $2$nd and $3$rd Rossiter mode, respectively, and switches to the $4$th Rossiter mode for $[L/D]=6$.

\begin{figure*}
\includegraphics[width=0.9\textwidth]{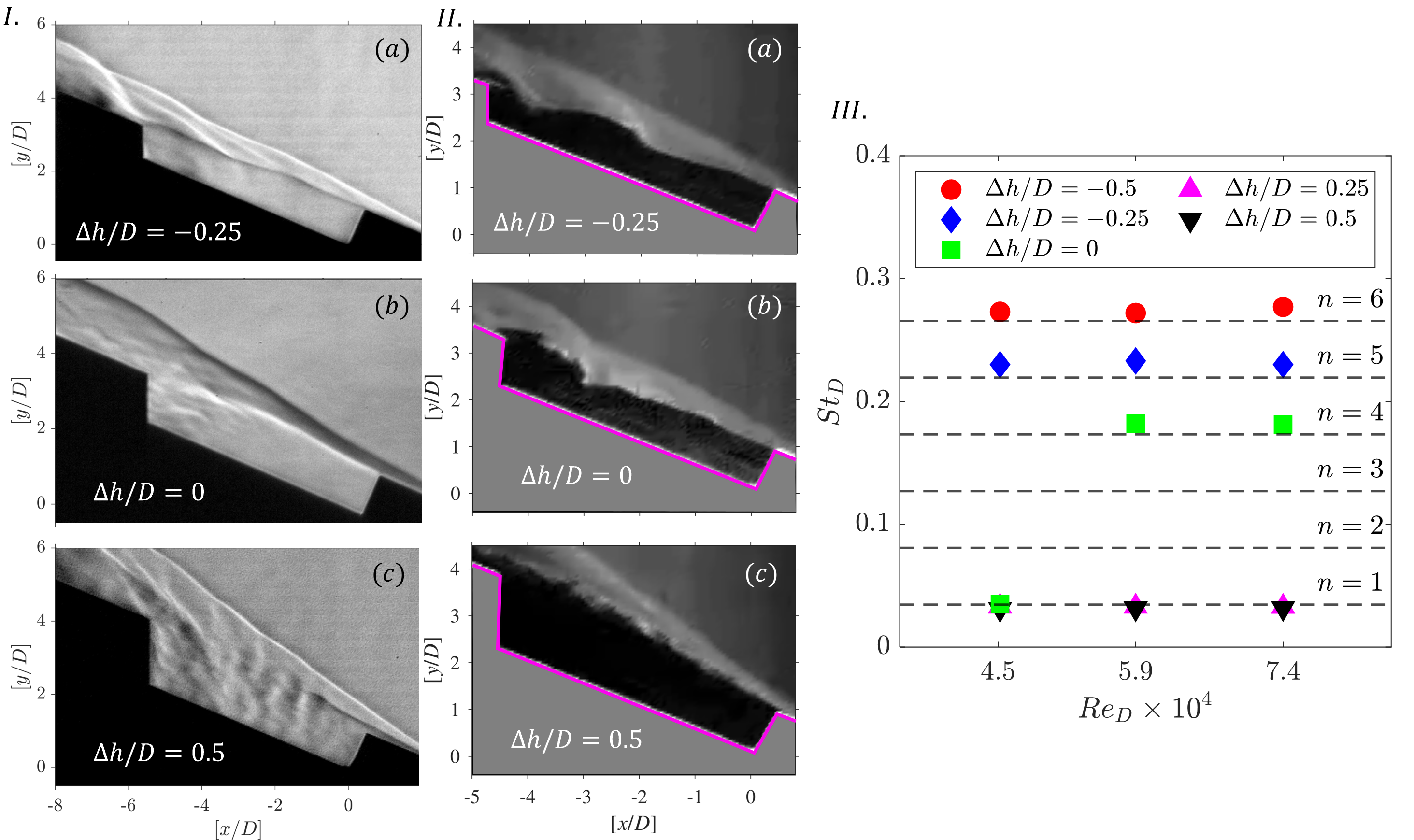} 
\caption{Instantaneous (I) high-resolution schlieren and (II) PLRS snapshots highlighting the overall flow field for (a) $[\Delta h/D] = -0.25$, (b) $[\Delta h/D] = 0$, and (c) $[\Delta h/D] = 0.5$ at $Re_D = 74000$. (III) Variation of $St_D$ with change in $Re_D$ across different cases of $[\Delta h/D]$ cases. The dashed horizontal lines correspond to Rossiter modes for $n=[1,2,..., 6]$ as calculated using Eq. \ref{eq:rossiter} and properties mentioned in Table \ref{tab:freestream_cond}. See the \href{https://youtu.be/31RkQYu9xLw}{supplementary} for viewing the corresponding video files.}
\label{fig:Model justification}
\end{figure*}

It is pertinent to note that on a similar line of observations as earlier analysis (Figure \ref{fig:sensor_PLRS_freq_LD} II-b), the dominant and secondary modes (Figure \ref{fig:pod_Re_D_ld} III-IV) clearly reveal a $Re_D$ dependence for the case of $[L/D]=6$. To highlight the flow structure difference, POD modes are displayed for the same three $Re_D$ cases presented in Figure \ref{fig:surf_freq_yd} II-IV. At $Re_D = 30000$ (Figure \ref{fig:pod_Re_D_ld} III-a), owing to mostly dominance of the laminar shear layer with minimal perturbation from the acoustic-hydrodynamic feedback loop as also corroborated by PLRS image (see Figure \ref{fig:PLRS_typical}a), the low amplitude flapping of the entire shear layer having $11\%$ of the flow energy is found to be the mode driving the flow. The second dominant mode (Figure \ref{fig:pod_Re_D_ld} IV-a), with $5\%$ of the energy, seems to be the harmonic mode of the weak shear layer flapping. So, it can be inferred that the identification of discrete frequency pertaining to the $1$st Rossiter mode (Figure \ref{fig:surf_freq_yd}-II) during the spectral analysis (see Sec. \ref{sec:quant L_d var}), corresponds to the entire shear layer lift-up due to the acoustic feedback loop. At $Re_D=51300$ (Figure \ref{fig:pod_Re_D_ld} III-b), the energy associated with the first mode nearly doubles to $23\%$, while still showing the shear layer lift-up as the dominant spatial feature, indicating amplified shear layer-acoustic wave interaction which is similar to flapping motion seen in leading-edge separated flows \citep{karthick2023}. However, the second dominant mode (Figure \ref{fig:pod_Re_D_ld} IV-b), accounting for $8\%$ of the energy, consists of alternating compression and rarefaction waves, a typical signature of the K-H instability. These two modes also establish the presence of multiple mode structures with significant energy content in the shear layer in line with earlier observations (Figure \ref{fig:surf_freq_yd}-III). At $Re_D=74000$ (Figure \ref{fig:pod_Re_D_ld} III-c), the POD distribution shows a qualitative switch, with K-H vortices having $\approx 24\%$ of the energy as the dominant mode, indicating the convection of instability waves driving the flow. Nevertheless, smaller scale of higher harmonics, having larger wave number structures with similar energy content ($20\%$) is perceived as the second dominant mode (Figure \ref{fig:pod_Re_D_ld} IV-c). Overall, the POD modes while increasing $Re_D$ for $[L/D]=6$, collectively demonstrate the strengthening of shear layer instability equivalent to flapping through amplified acoustic-hydrodynamic coupling, followed by switching to K-H instability and eventually introduce finer-scale transitional structures.

\subsection{Effect of Cavity's Excess Rear Face Height, $[\Delta h/D]$}

\begin{figure*}
	\includegraphics[width=0.82\textwidth]{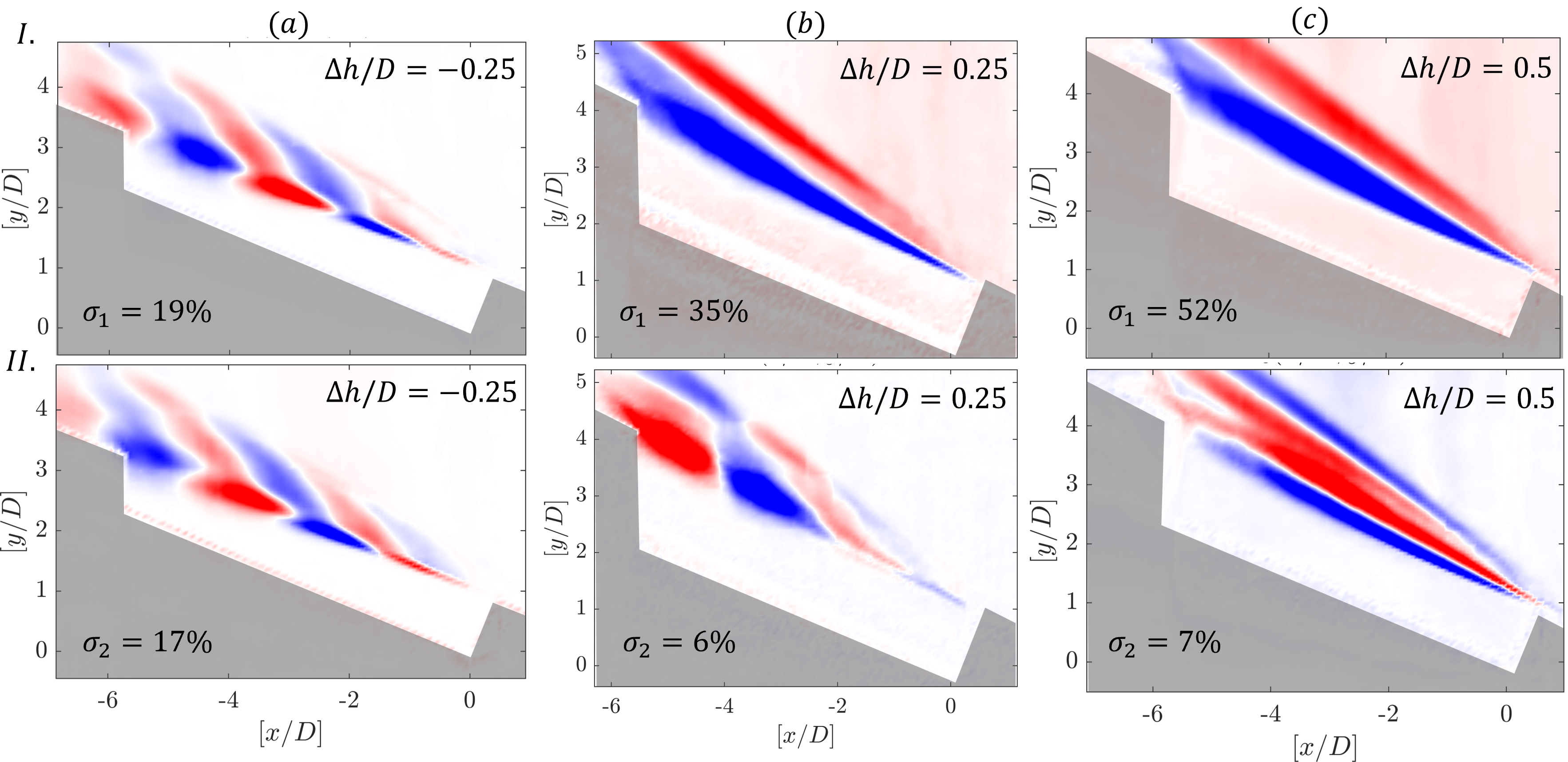}
	\caption{Spatial structure corresponding to the dominant (I) first and (II) second mode obtained through POD analysis for $[L/D]=6$ at different $[\Delta h/D]$ of (a) $-0.25$, (b) $0.25$, and (c) $0.5$. $\sigma$ represents the respective model energy content.}
    \label{fig:pod_hd_ld6}
\end{figure*}

Experiments are conducted for $[L/D]=6$ at specific $Re_D = [45000, 59000, 74000]$ (ref. Figure \ref{fig:Model justification}-III), across different $[\Delta h/D]$s ranging from $-0.5$ to $0.5$ in an interval of $0.25$. PLRS is performed for all the cases, and the temporal variation in light intensity is extracted at the intersection point of the shear layer and the cavity center line. Afterwards, the Fast Fourier Transform (FFT) is performed to extract the spectral content, and the dominant frequency for each case is noted. The variation of the dominant frequency in terms of $St_D$ is obtained and compared with different Rossiter modes as shown in Figure \ref{fig:Model justification}-III. It is evident that there is a change in flow interaction mechanism for $[\Delta h/D]=0$ as the dominant mode shifts with $Re_D$, which has been discussed in the previous section. For the negative cases of $[\Delta h/D]$, the shear layer is comprised of convective K-H vortices (see Figure \ref{fig:Model justification} I,II-a for $[\Delta h/D]=-0.25$), thereby indicating the same mechanism for $[\Delta h/D]=[-0.25, -0.5]$. Further, in the case of positive cases of $[\Delta h/D]$, the dominant mode number is found to be invariant across $Re_D$. Nevertheless, upon analysis through POD, it is observed that for $[\Delta h/D]=0.25$ (see Figure \ref{fig:pod_hd_ld6}b), the energy content spans both mode numbers $n=1$ and $n=4$, with a significant contribution from the former. However, for $[\Delta h/D]=0.5$, no spatial structure is seen convecting along the shear layer which is affirmed by the instantaneous images (see Figure \ref{fig:Model justification} I,II-c and corresponding videos in the \href{https://youtu.be/31RkQYu9xLw}{supplementary}). As the present study attempts to investigate the flow evolution across different standalone flow scenarios, we shall concentrate on cases having one type of dominant spatial mode ($[\Delta h/D] = [-0.25,0.5]$) while varying $[\Delta h/D]$.

\begin{figure*}
	\includegraphics[width=\textwidth]{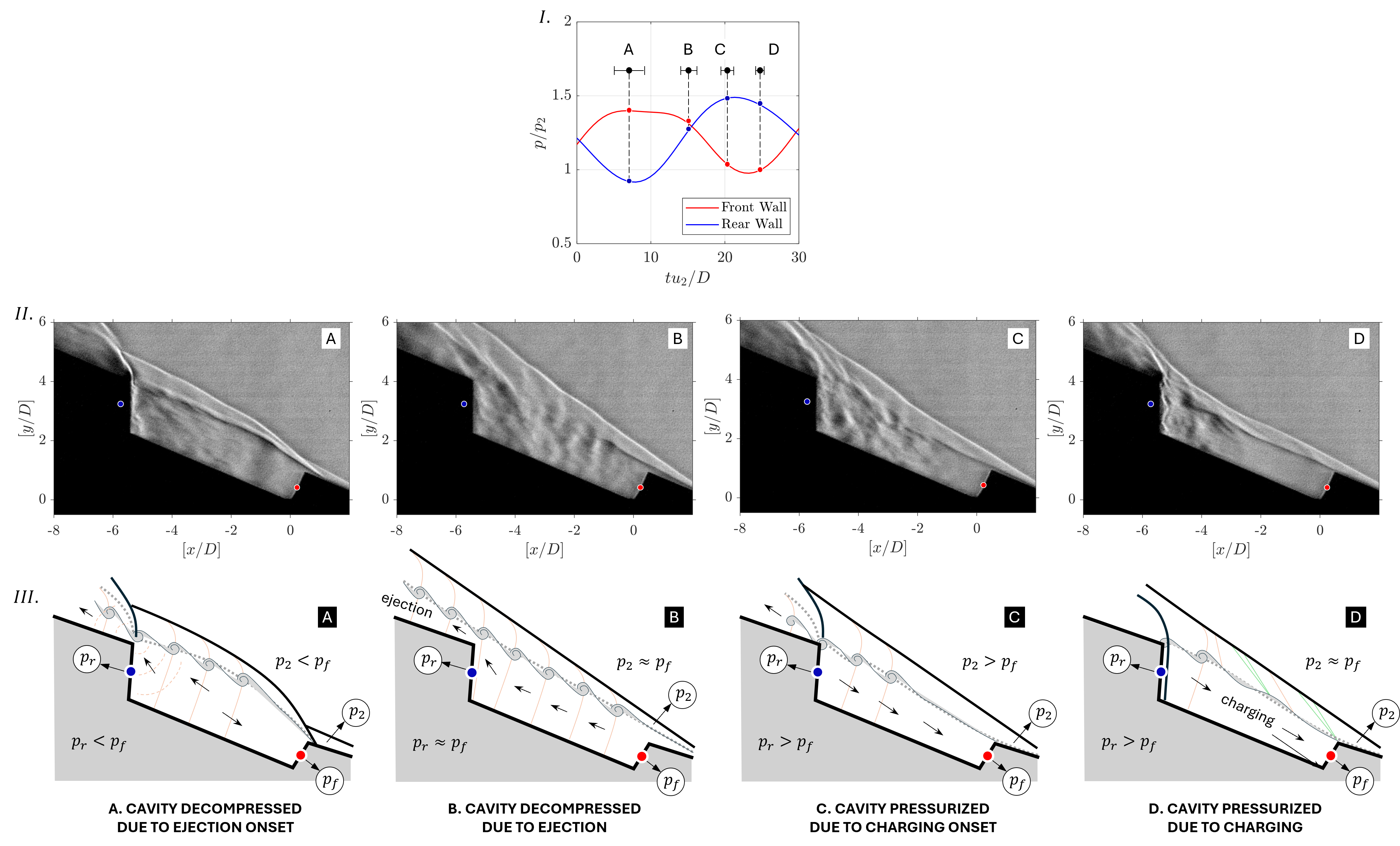}
	\caption{\label{fig:pulsation_mech} (I) The pressure trace obtained from the front ($p_f$) and rear pressure sensor ($p_r$) for one breathing cycle of flapping motion depicted for the case of $[\Delta h/D]=0.5$ at $Re_D=74000$. The flapping cycle is split into four representative frames (A-D) that are presented in terms of (II) instantaneous schlieren snapshots, and (III) schematic explaining the underlying flow scenario. The expected pressure at $p_f$ and $p_r$ during each of the frames are marked in the Figure \ref{fig:pulsation_mech}-I. In the schematic, flow features such as leading-edge shock, impingement shock, separation shock are marked in dark black color, with weak shocklets and acoustic waves marked in dashed orange color, and expansion fan in green color.}
\end{figure*}

The instantaneous schlieren images are captured at a higher frequency (sampling rate of $52$ kHz) only for the cases of $[\Delta h/D] = 0.5$ and $Re_D=74000$, to better understand the underlying mechanism driving the unsteadiness. Simultaneously, pressure responses at the front and trailing edges of the cavity are also acquired to correlate the events with the schlieren snapshots. As a whole, K-H vortices convecting along the shear layer without lift-up are observed for $[\Delta h/D] = -0.25$, similar to characteristics of a regular open cavity configuration. However, the entire shear layer and shock system is seen to be flapping for the case of $[\Delta h/D] = 0.5$. For one cycle of flapping, the typical pressure response obtained from the front and rear sensors for $[\Delta h/D] = 0.5$ is shown in Figure \ref{fig:pulsation_mech}-I. The relevant schlieren frames (A-D) necessary to explain one flapping cycle are shown in Figure \ref{fig:pulsation_mech}-II, along with the associated pressure magnitudes marked in Figure \ref{fig:pulsation_mech}-I. Also, for better clarity, the corresponding simplified schematic depicting the wave diagram in line with the instantaneous schlieren snippet is shown in Figure \ref{fig:pulsation_mech}-III. The pressure traces, in conjunction with the schlieren frames, indicate that the $[\Delta h/D] = 0.5$ case undergoes a distinct breathing cycle driven by alternating under- and over-expansion. To isolate the events, it is split into four frames, presented in Figure \ref{fig:pulsation_mech}-II and Figure \ref{fig:pulsation_mech}-III. Frame A (Figure \ref{fig:pulsation_mech}-II) shows the flowfield close to the onset of depressurization of the cavity demarcated by shear layer lift-up near the leading edge. The pressure imbalance, with higher pressure in the vicinity of the front wall than the rear wall ($p_f>p_r$), has resulted in a strong upward motion of the shear layer and local over-expansion near the lip of the cavity leading edge ($p_f>p_2$). The cavity trailing zone appears to be filled with fine ripples and tiny compressed spots, representing acoustic waves and weak shocklets. As the cavity begins to transition from its previously compressed state, the rear region relaxes through weaker pressure pulses and acoustic waves. Also, the increased front-face pressure ($p_f$) at the cavity lip results in upstream motion of the shear layer and separation shock, which is evident in Frame B of Figure \ref{fig:pulsation_mech}-II. 

\begin{figure*}
	\includegraphics[width=0.9\textwidth]{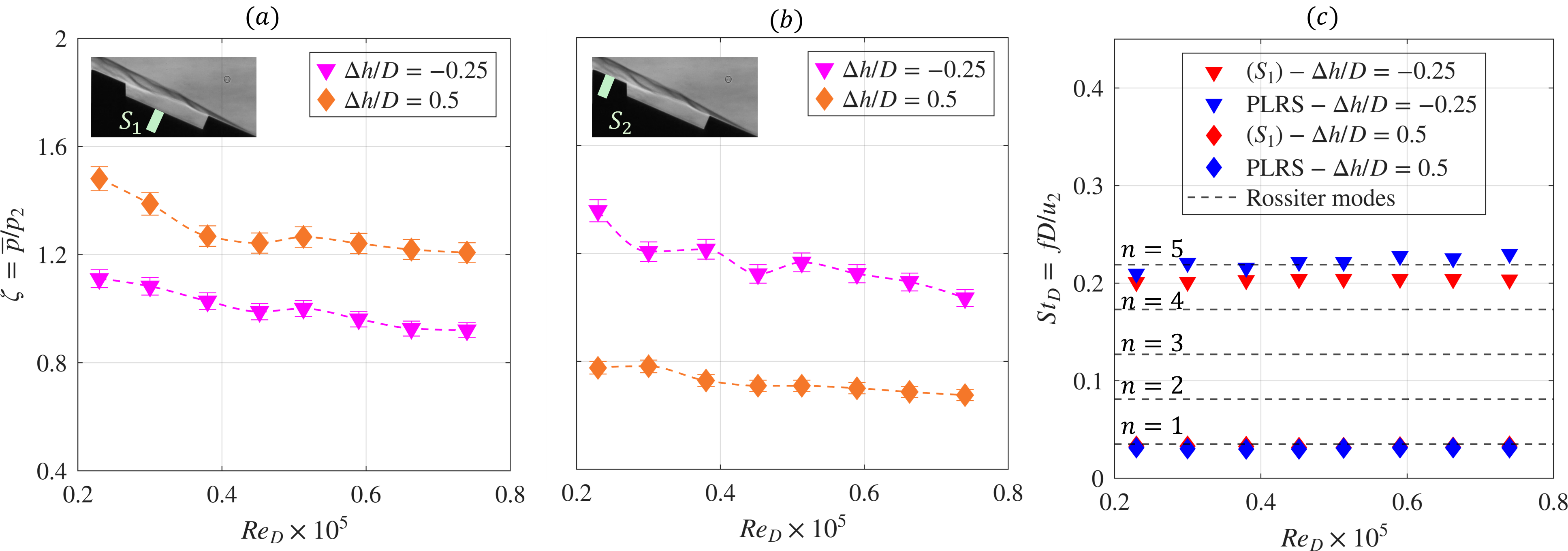}
	\caption{Non-dimensional mean pressure variation at the (a) cavity floor center, $S_1$ and (b) cavity downstream, $S_2$ location with change in $[\Delta h/D]$ and $Re_D$ at $[L/D] = 6$. (c) Comparison of dominant frequency obtained from the sensor $(S_1)$ and PLRS image based processing with change in $Re_D$ for $[\Delta h/D] = [-0.25, 0.5]$.}
    \label{fig:p_hbyD_effect}
\end{figure*}

The venting process continues from Frame A to B, where Frame B represents a flow situation where the separation shock reaches its extreme extent, resulting in pressure equalization of the cavity interior with the freestream flow ($p_f\approx p_r \approx p_2$). With the shear layer no longer supported by excess front pressure, the external supersonic flow begins to push the shock and shear layer downward. The downward motion is initiated by admitting a small amount of high-speed flow back into the cavity through the first set of weak compression waves, as can be clearly seen in Frame B of Figure \ref{fig:pulsation_mech}-II. With further leaning of the shear layer towards the cavity, high-speed outer flow entrains into the cavity before reaching a stable configuration where the shear layer strikes the rear wall, resulting in a strong impingement shock (see Frame C in Figure \ref{fig:pulsation_mech}-II), which is also the onset of the recompression phase of the cavity. The schlieren snippet in Frame C reveals a high-contrast density-gradient representation of shocklets near the trailing edge, resulting in a pressure rise in $p_r$. In the mean time, also the forcing from the freestream flow pushes the shear layer fully downward and locking at its maximum inward curvature, with the presence of a centered expansion fan around the leading edge lip of the cavity, as seen in Frame D of Figure \ref{fig:pulsation_mech}-III. Also, the impingement shock has grown into a steep, anchored shock with its foot on the cavity floor, indicating that high-momentum outer flow is being injected into the cavity, resulting in a continuous pressurization, which, at a later time, pushes the shear layer upward, thereby completing the cycle. Furthermore, the mean pressure is measured across several runs and oscillation cycles at the $p_f$ and $p_r$ locations, which results in a pressure gradient of $dp/dx = \left(\left(\overline {p_f}-\overline {p_r}\right)/p_2\right)/(L/D)$ of $0.0323$, indicating the intermittent excess pressure at the front face causing the upward movement of the shear layer near the leading edge that drives the overall unsteadiness.

The instantaneous schlieren snapshots for $[\Delta h/D]=-0.25$ at $Re_D=74000$ (see Figure \ref{fig:Model justification} I,II-a), reveal a shear layer with the presence of less intense K-H vortices and the associated weak shocklets within the shear layer. The shear layer impingement consistently overshoots the rear lip, reattaching downstream of the trailing edge with embedded K-H vortices, confirming that the shear layer is mainly governed by the convective instability, unlike the pressure build-up for $[\Delta h/D]=0.5$. Also, due to the shear layer impingement downstream of the rear face, only weak perturbations transmit into the cavity, thereby helping in prohibiting the triggering of transition as seen in $[\Delta h/D]=0$ (see Figure \ref{fig:Model justification} I,II-b). Also, the shear layer seems to support smaller-wavelength vortical structures owing to a longer shear layer path and a weaker pressure-feedback loop. The pressure gradient across the leading and trailing edges measured from the $p_f$ and $p_r$ sensors is calculated to be $-0.0033$, one order of magnitude lower than the case of $[\Delta h/D]=0.5$. The relatively weaker K-H vortices convecting downstream shed weaker compression waves, which are felt mainly by the $p_r$ sensor. Since the cavity-driven resonance is weaker, the overall cavity pressure remains low, and the downstream convective vortices naturally produces marginally higher $p_r$ than $p_f$.

\begin{figure*}	               

\includegraphics[width=0.85\textwidth]{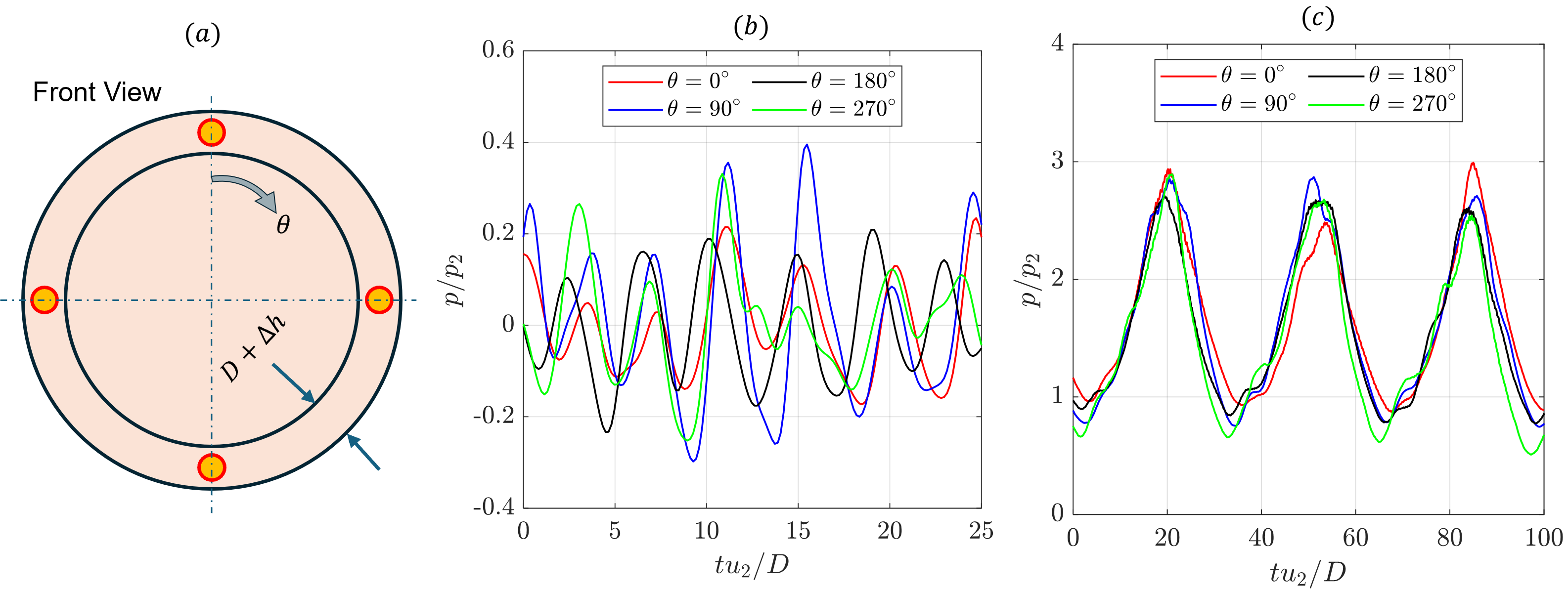}
	\caption{(a) The locations of the sensor's placement along the azimuthal direction on the rear face of the cavity. The pressure responses obtained from these sensors are plotted for the case of (b)$[\Delta h/D]=-0.25$, and (c) $[\Delta h/D]=0.5$, depicting no clear corelation for $[\Delta h/D]=-0.25$, and coherent signals for $[\Delta h/D]=0.5$. }
    \label{fig:azimuthal_press}
\end{figure*}

The normalized mean pressure inside the cavity ($\zeta = \overline{p}/p_2$) during the test duration is measured from the response of $S_1$. The value of $\zeta$ for $[\Delta h/D] = 0.5$ is found to be higher than that of the $[\Delta h/D] = -0.25$, irrespective of the $Re_D$ (see Figure \ref{fig:p_hbyD_effect}a). The significant increase in $\zeta$ corresponds to distinct patterns of flow unsteadiness observed during the schlieren imaging reported previously. The case of $[\Delta h/D] = 0.5$ pertains to a situation where the entire shock and shear layer system flap severely, thereby imparting a higher cyclic load on the cavity, similar to leading-edge-separated flows \citep{karthick2023}. However, for $[\Delta h/D] = -0.25$, the separated shear layer is perturbed by the K-H instability structures spanning the entire cavity (see Figure \ref{fig:Model justification} I,II-a), resulting in a lesser pressure loading. Comparing Figure \ref{fig:p_hbyD_effect}a,b, the flapping mode is found exerting higher pressure inside the cavity ($S_1$), while the K-H mode on the downstream cavity location ($S_2$). The impingement of the shear layer downstream of the trailing edge for $[\Delta h/D] = -0.25$ causes higher pressure for the cavity location $S_2$. Also, owing to the cavity breathing motion seen for $[\Delta h/D] = 0.5$, the impingement shock strength varies intermittently, to ensure flow relaxation, leading to a lesser pressure rise at the downstream location. Nevertheless, the trend of decrement in $\zeta$ with increase in $Re_D$ is common for both the inside and downstream of the cavity cases (see Figure \ref{fig:p_hbyD_effect}) .

The dominant frequency obtained from the sensor ($S_1$) and the PLRS-based FFT analysis for two $ [\Delta h/D]$ cases is shown in Figure \ref{fig:p_hbyD_effect}c. For $[\Delta h/D] = 0.5$, the dominant $St_D$ is quantified as $0.036$, which is the same as observed during the post-processing of the high-repetition-rate PLRS imaging. Also, this non-dimensional frequency matches with the $1$st Rossiter mode obtained from Eq. \ref{eq:rossiter} for $[L/D] = 6$. Similarly, for $[\Delta h/D] = -0.25$, the dominant non-dimensional frequency is found to be $[fD/u_2] = 0.2$, which has a slight deviation from the values obtained from the PLRS-based imaging ($[fD/u_2] = 0.21$). Moreover, these values are closer to Rossiter's $5$th mode ($[fD/u_2] = 0.217$). For both the cases of $[\Delta h/D] = -0.25, 0.5$, the dominant frequency is found to be invariant of $Re_D$. 

With the same analogy as $[L/D]$ variation, 600 PLRS frames during the steady test duration are subjected to POD towards extracting the dominant spatial mode driving the flow. The identified spatial modes are displayed in Figure \ref{fig:pod_hd_ld6}a,c for $[\Delta h/D]=-0.25, 0.5$, respectively. In line with observation of weaker K-H vortices along the shear layer in the schlieren snapshots, the dominant mode is also found to have $19\%$ energy (see Figure \ref{fig:pod_hd_ld6} I-a) in comparison to $24\%$ energy for the case of $[\Delta h/D]=0$ (see Figure \ref{fig:pod_Re_D_ld} III-c). Both the primary and secondary dominant modes for $[\Delta h/D]=-0.25$ establish the presence of K-H vortices in the shear layer through alternating red and blue patches with a phase shift for the secondary mode. Also, the number of vortices spanning the cavity reaffirms the match between the dominant frequency and the $5$th Rossiter mode. Together, the spatial modes indicate the significance of convective-dominated shear layer towards driving the overall flow. In contrast, the first dominant mode for $[\Delta h/D]=0.5$ constitutes $52\%$ of the total flow energy (see Figure \ref{fig:pod_hd_ld6} I-c), signifying the large-scale pressure-driven motion resulting in bulk vertical displacement of the entire shear layer. A similar flow pattern has also been reported in leading-edge separated flows \citep{karthick2023}. The strong intensity of the primary mode indicates that the cavity behaves like a single low-frequency oscillating fluid volume. The secondary mode (see Figure \ref{fig:pod_hd_ld6} II-c) seems to be a complementary phase of the coherent vertical displacement of the shear layer. The quantitative signature of the breathing mechanism matches the frequency associated with the $1$st Rossiter mode.

\begin{figure*}
	\includegraphics[width=0.8\textwidth]{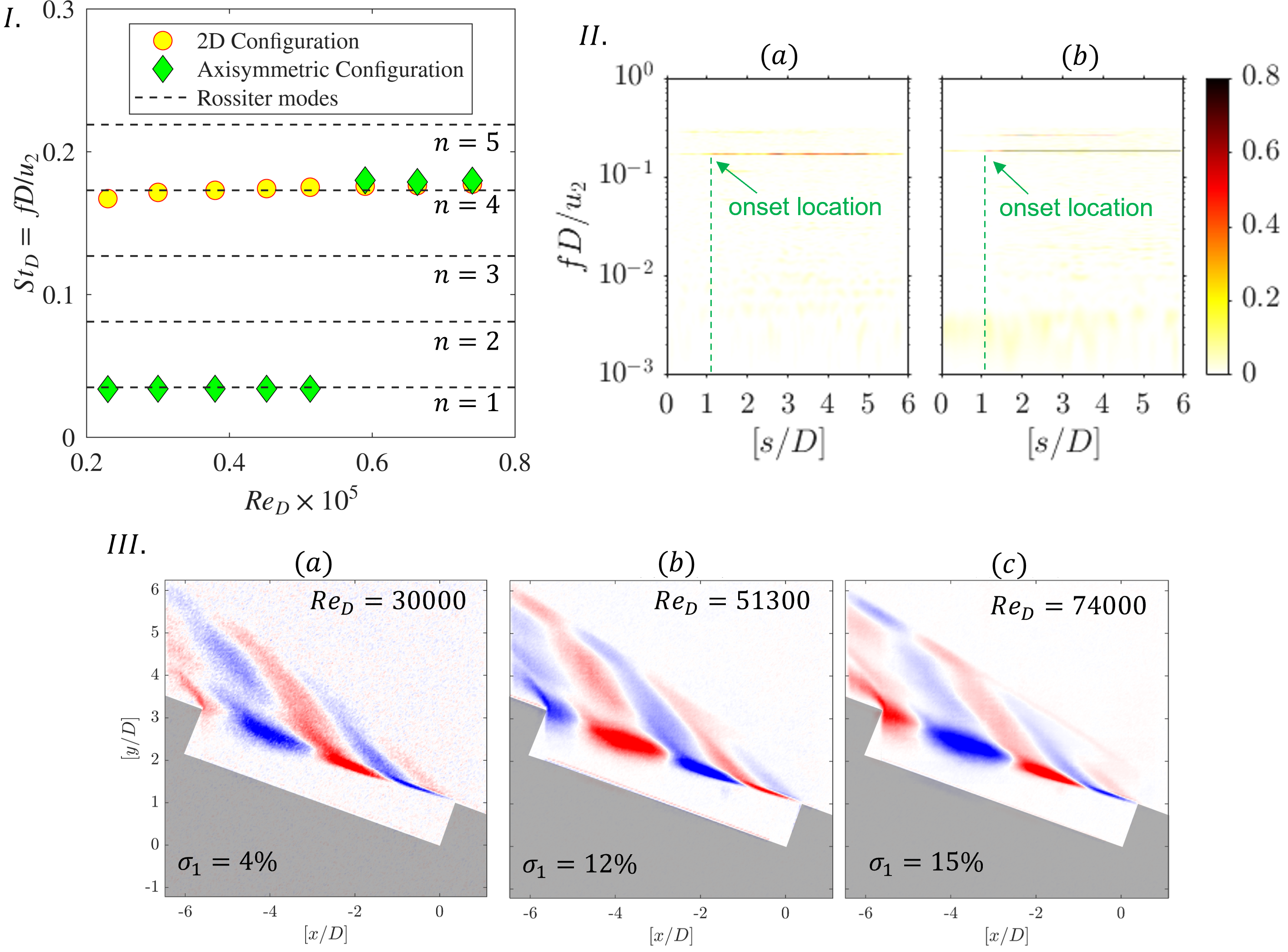}
	\caption{(I) Comparison of the dominant frequency obtained from the PLRS image based processing for the two-dimensional and axisymmetric cavity configuration for $[L/D] = 6$. The dashed black lines represent the Rossiter modes obtained using the semi-empirical relation as mentioned in Eq. \ref{eq:rossiter}. (II) $s-f$ diagram generated for the case of two-dimensional cavity configuration having $[L/D] =6$ at (a) $Re_D = 23000$, and (b) $Re_D = 74000$ where the profile is generated along the shear layer. A representative line along which $s-f$ plot is generated, is marked in dashed green color in Figure \ref{fig:PLRS_typical}b. (III) Dominant spatial structure obtained through POD analysis for two-dimensional cavity having $[L/D]=6$ at different $Re_D$ of (a) $30000$, (b) $51300$, and (c) $74000$. See the \href{https://youtu.be/rZRJ0eobYKQ}{supplementary} for viewing the corresponding video files.}
    \label{fig:2D_axisym}
\end{figure*}

With variation of $[\Delta h/D]$, two different spatial flow patterns dominating the flow and leading to unsteadiness are evident. In order to determine the azimuthal nature of these oscillations, four pressure probes equally spaced are placed along the circumferential direction (see Figure \ref{fig:azimuthal_press}a). Experiments are performed for the two representative cases of $[\Delta h/D] = [-0.25,0.5]$, at a freestream $Re_D=74000$. The responses obtained from the sensors in the case of $[\Delta h/D] =-0.25$ are shown in Figure \ref{fig:azimuthal_press}b, from which a clear difference in amplitude and phase among all the azimuthal locations is evident. On the other hand, for the case of $[\Delta h/D] = 0.5$, the pressure fluctuations across all the azimuthal locations exhibit near identical amplitudes without any significant phase delay (see Figure \ref{fig:azimuthal_press}c), confirming that the flapping mode is axisymmetric.

To further quantify the azimuthal behavior, modal decomposition from the acquired responses is carried out for the respective dominant Rossiter frequency ($f_0$) using the approach described by Cohen and Wygnanski (1987)\citep{cohen1987} in Sec. 2.2. This allows the separation of the amplitude at $f_0$ into axisymmetric (denoted by $m=0$) and the first azimuthal component (denoted by $m=\pm 1$), which are also representative of forward- and backward-rotating modes.  The modal separation for $[\Delta h/D] = -0.25$ (K-H mode) resulted in amplitudes of $m=0,\ +1$, and $-1$ as $0.053,\ 0.046$, and $0.031$, respectively, which are small and comparable in magnitude without any preferential direction of propagation. It is also to be noted that any specific corelation could not be established among the sensors which is in agreement with findings reported by Creighton and Hillier (2007) \citep{creighton2007}. Similarly, for the $[\Delta h/D] = 0.5$ (flapping mode), the modal separation results in amplitudes of $m=0,\ +1$, and $-1$ as $0.6,\ 0.044$, and $0.038$, respectively. The dominance of the axisymmetric mode in comparison to the helical modes establishes the azimuthal uniformity of the flow features for the case of $[\Delta h/D] = 0.5$. 

\subsection{Flow Disparity between Two-Dimensional and Axisymmetric Cavities}

The following section highlights the differences in flow evolution between axisymmetric and two-dimensional cavities across the $Re_D$ range for $[L/D]=6$. In this regard, a two-dimensional cavity configuration having $w/D=10$ is prepared with a wedge angle of $20.5^\circ$, which ensures the flow properties exposed by the cavity remain the same as Table \ref{tab:freestream_cond}. The $L_F$ is also kept the same, so that the incoming boundary layer thickness at the leading edge also remains consistent. The temporal variation of light intensity obtained through PLRS at the shear layer midpoint is evaluated, and further FFT is performed to identify the dominant frequency. The comparison of dominant frequency across the $Re_D$ cases for both the axisymmetric and two-dimensional cavity configurations is depicted in Figure \ref{fig:2D_axisym}-I. It is clearly evident that the dominant frequency matches with the $4$th Rossiter mode for all the $Re_D$s in the 2D cavity, demonstrating the $Re_D$ independence of the dominant frequency, in agreement with earlier studies \citep{chung2020, Heller1971}, in contrast to the mode switching behavior from $1$st to $4$th Rossiter mode observed for the axisymmetric cavities.


Further, in order to have a quantitative comparison on the shear layer perturbation location for the two-dimensional configuration, $s-f$ plots are generated as shown in Figure \ref{fig:2D_axisym}-II. Much like the dominant frequency, the perturbation onset along the shear layer is also found to be constant ($s/D = 1.05$), which demarcates the coherence of the separated shear layer across the $Re_D$ cases. This is again in contrast to the axisymmetric cavity case, where upstream movement of the perturbation location is seen with increases in $Re_D$ (see Figure \ref{fig:xf_LD6}), which further results in shear layer breakdown owing to longer propagation length from the perturbed location to the trailing edge. Moreover, the instantaneous PLRS images are subjugated through the POD technique to arrive at the dominant mode, which is presented in Figure \ref{fig:2D_axisym}-III. All the $Re_D$ cases display convecting K-H vortices of identical characteristic wavelength that extend from the leading edge to the trailing edge, thereby establishing the invariant of the dominant $St_D$ in line with PLRS-based spectral analysis (see Figure \ref{fig:2D_axisym}-I). With increases in $Re_D$, the amplification of instability wave occurs (compare Figure \ref{fig:2D_axisym} III-a \& \ref{fig:2D_axisym} III-c) without altering the underlying topology of the coherent structures, thereby confirming that irrespective of the $Re_D$, the dominant dynamics are driven by the same Rossiter-type shear layer instability. Nevertheless, the axisymmetric cavity at lower $Re_D$s (see Figure \ref{fig:pod_Re_D_ld} III-a) is found to be driven by the shear layer lift-up (Rossiter $1$st mode), which intensifies at intermediate $Re_D$s (see Figure \ref{fig:pod_Re_D_ld} III-b) along with significant contribution from the Rossiter $4$th mode (see Figure \ref{fig:pod_Re_D_ld} IV-b). Finally, at higher $Re_D$s (see Figure \ref{fig:pod_Re_D_ld} III-c), K-H vortices, along with fine-scale turbulent structures resulted out of shear layer transition to turbulence, is seen to be dominating the overall flow field.

\section{Summary and Conclusions}

Experiments to understand the shear layer characteristics and identify the source of unsteadiness during hypersonic flow over axisymmetric cavities mounted on a cone are performed using the Hypersonic Ludwieg Tunnel (HLT) located at Technion. Various cavity geometries with different aspect ratios ($[L/D] = [2,\ 4,\ 6]$) are exposed to hypersonic flow having a freestream Mach number $M_\infty=6.0$, at a wide range of Reynolds numbers ($23000\leq Re_D \leq 74000$). Additionally, the effect of varying the shear layer reattachment location through normalized excess height of the rear face ($[\Delta h/D] = [-0.5,-0.25,0,0.25,0.5]$) is also assessed for $[L/D]=6$. High-resolution schlieren snapshots reveal the overall flow features, including the leading-edge shock, the separated shear layer, K-H vortices, and the reattachment shock. Similarly, high-repetition-rate PLRS imaging is used to construct space-time ($x-t$) and space-frequency ($x-f$) plots to establish the unsteadiness level for different cases. Along with the qualitative interpretation, quantitative estimates are obtained using pressure probes at different locations, enabling identification of the mean pressure variation and the dominant frequency through FFT.

\begin{itemize}

\item \textbf{Effects of length to depth ratio ($[L/D]$)}: Increasing $[L/D]$ fundamentally alters the shear layer dynamics through amplifying instability and strengthening acoustic-hydrodynamic coupling. For a relatively short cavity ($[L/D] =2$), the schlieren and PLRS snapshots reveal a predominantly laminar shear layer with lesser intense K-H vortices convecting at a frequency that matches the $2$nd Rossiter mode, which is also evident from the POD analysis that shows one set of alternate colored-lobes that span the entire cavity length. The pressure response within and downstream of the cavity is nearly identical, indicating a stable recirculation region. For $[L/D]=4$, visible K-H vortices with intensified fluctuation levels and a reattachment shock with increase in $Re_D$ are observed from the qualitative imaging, which, upon analysis, shows Reynolds number independent behavior for $St_D$ that matches the $3$rd Rossiter mode, along with a modest pressure variation between the cavity center and downstream location. For a relatively longer cavity ($[L/D]=6$), the shear layer becomes highly receptive, with the associated perturbation location shifting upstream as $Re_D$ increases, resulting in transition to turbulence owing to longer traverse length for the shear layer accompanied by amplified K-H instabilities. At intermediate $Re_D$s, both, the flapping ($n=1$), dominating the low $Re_D$ range, and the K-H ($n=4$), dominating the high $Re_D$ range, coexist. Therefore, at this range, the dominant spectral peak detected by the sensor, might depend on its location. It should be noted that the K-H mode is observed to prevail throughout the entire $Re_D$ range in the two-dimensional cavity case under the same flow conditions. The pressure measurement at the cavity center approaches the freestream value as the shear layer transitions to turbulence. The latter also results in $\approx20\%$ lower pressure at the downstream reattachment location. For $[L/D]=6$, POD analysis suggests that for the shear layer, the lift-up is similar to the flapping behavior at lower $Re_D$s, and that the K-H vortices of different wave number dominate the flow at higher $Re_D$s.

\item \textbf{Effects of excess rear face height to depth ratio ($[\Delta h/D]$)}: For the negative excess rear face height to depth ratio ($[\Delta h/D]=-0.25$), schlieren images display K-H vortices convecting downstream without appreciable lift-up, and the reattachment shock constantly overshoots the trailing edge, resulting in a relatively weakly forced shear layer scenario and a minimal pressure gradient along the cavity. POD analysis also establishes the dominance of K-H vortices with associated frequency matching the $5$th Rossiter mode. In contrast, for the positive excess rear face height to depth ratio ($[\Delta h/D]=0.5$), schlieren revealed a pronounced flapping of the shear layer, for which the $St_D$ matches the $1$st Rossiter mode across all the $Re_D$ cases. The flapping mode is similar to the shock-shear layer oscillation observed in leading-edge separated flows. One cycle of the cavity breathing motion is prompted by the alternate under- and over-expansion process. This involves the shear lift-up together with upstream movement, depressurization of the cavity, leading to the maximum extent of the shock-shear layer system. The latter events are followed by the cavity pressurization induced by the strong impingement shock having a foot on the cavity floor, as well as the presence of shocklets near the trailing edge. POD for $[\Delta h/D]=0.5$ also indicates the presence of a single strongly coupled resonant structure spanning across the cavity length as the dominant one. For the positive excess rear face height to depth ratio, strong flapping results in higher pressure within the cavity, whereas for the negative excess rear face height to depth ratio, downstream impingement of the shear layer leads to higher pressure at the post-cavity location. Azimuthal pressure variation on the cavity rear face and further post-processing via mode separation indicate the axisymmetric nature of the flapping mode, whereas K-H vortices did not show significant correlated behaviour.

\end{itemize}

\section*{Supplementary material}
The manuscript contains the following videos: 
\begin{enumerate}
\item \href{https://youtu.be/IWpR-vujNLw}{Video (click to play)} of the high-repetition-rate PLRS and schlieren imaging highlighting the effect of $[L/D]$ at $[\Delta h/D]=0, Re_D=74000$.
\item \href{https://youtu.be/VRJ5KJEjSIE}{Video (click to play)} of the high-repetition-rate PLRS and schlieren imaging towards explaining the effect of $Re_D$ for $[L/D]=6, [\Delta h/D]=0$.
\item \href{https://youtu.be/31RkQYu9xLw}{Video (click to play)} of the high-repetition-rate PLRS and schlieren imaging portraying the effect of normalized excess rear face height ($[\Delta h/D]$ at $Re_D=74000$ for $[L/D]=6$. 
\item \href{https://youtu.be/rZRJ0eobYKQ}{Video (click to play)} presenting the differences in flow evolution between 2D and axisymmetric cavities through PLRS imaging for $[L/D]=6$ at $Re_D=74000$.
\end{enumerate}

\section*{Acknowledgments} 
The authors would like to acknowledge the help of the lab members Michael Dunaevsky, Oleg Kan, Nadav Shefer, David Naftali, and Efim Shulman towards conducting the experiments in the Hypersonic Ludwieg Tunnel (HLT) at Technion Wind Tunnel Complex.



\section*{References}
\vspace{0.1cm}
\nocite{*}
\bibliography{references}

\end{document}